\documentclass[fleqn,usenatbib]{mnras}
\usepackage{newtxtext,newtxmath}
\usepackage[T1]{fontenc}

\DeclareRobustCommand{\DA}[3]{#2}
\let\DAthebibliography\thebibliography
\def\thebibliography{\DeclareRobustCommand{\DA}[3]{##3}\DAthebibliography}
\DeclareRobustCommand{\daname}[3]{#2}
\let\dathebibliography\thebibliography
\def\thebibliography{\DeclareRobustCommand{\daname}[3]{##3}\dathebibliography}

\usepackage{graphicx}	
\usepackage{amsmath}	
\usepackage[dvipsnames]{xcolor} 

\newcommand{\bagpipes}{{\sc bagpipes}}

\title[Stellar and Dust Emission Fitting]{On the Simultaneous Modelling of Dust and Stellar Populations  for Interpretation of Galaxy Properties}

\author[G. T. Jones et al.]{
Gareth T. Jones,$^{1}$\thanks{E-mail: g.jones.6@warwick.ac.uk}
Elizabeth R. Stanway\,$^{1}$
and
A. C. Carnall\,$^2$
\\
$^{1}$ Department of Physics, University of Warwick, Gibbet Hill Road, Coventry, CV4 7AL, UK\\
$^2$ Scottish Universities Physics Alliance, Institute for Astronomy, University of Edinburgh, Royal Observatory, Edinburgh, EH9 3HJ, UK
}

\date{Accepted XXX. Received YYY; in original form ZZZ}

\pubyear{2022}

\begin{document}
\label{firstpage}
\pagerange{\pageref{firstpage}--\pageref{lastpage}}
\maketitle

\begin{abstract}
The physical properties of galaxies are encoded within their spectral energy distribution and require comparison with models to be extracted. These models must contain a synthetic stellar population and, where infrared data is to be used, also consider prescriptions for energy reprocessing and re-emission by dust. While many such models have been constructed, there are few analyses of the impact of stellar population model choice on derived dust parameters, or vice versa. Here we apply a simple framework to compare the impact of these choices, combining three commonly-used stellar population synthesis models and three dust emission models. We compare fits to the ultraviolet to far-infrared spectral energy distributions of a validation sample of infrared-luminous galaxies. We find that including different physics, such as binary stellar evolution, in the stellar synthesis model can introduce biases and uncertainties in the derived parameters of the dust and stellar emission models, largely due to differences in the far-ultraviolet emission available for reprocessing. This may help to reconcile the discrepancy between the cosmic star formation rate and stellar mass density histories. Notably the inclusion of a dusty stellar birth cloud component in the dust emission model provides more flexibility in accommodating the stellar population model, as its reemission is highly sensitive to the ultraviolet radiation field spectrum and density. Binary populations favour a longer birth cloud dissipation timescale than is found when assuming only single star population synthesis.

\end{abstract}

\begin{keywords}
galaxies:stellar content -- dust, extinction -- techniques: photometric
\end{keywords}



\section{Introduction}

Stellar population synthesis (SPS) models have been developed since the late 1960s \citep{1968ApJ...151..547T} and used to interpret the optical component of galaxy spectral energy distributions (SEDs). They exploit stellar evolution theory to model a range of possible stellar populations and simulate their photometry, which can then be matched to observations \citep[for an extensive review, see][]{2013ARA&A..51..393C}. SPS models have constantly evolved to incorporate developing knowledge of stellar physics, including the physics of stellar atmospheres \citep[e.g.][]{2011A&A...532A..95F}. A more recent development has been to include binary evolution pathways for stars in addition to those arising from isolated single star evolution \citep[e.g.][]{2002ApJ...572..407B, 2008ApJS..174..223B, 2017NatCo...814906S, 2009MNRAS.400.1019E,2017PASA...34...58E, 2018MNRAS.481.1908K}. Around 70 per cent of massive stars are believed to interact as part of a binary over the course of their evolution \citep[e.g.][]{2009ApJ...692..618M, 2012MNRAS.424.1925C, 2012Sci...337..444S, 2013A&A...550A.107S, 2017ApJS..230...15M} and these can substantially affect the interpretation of young stellar populations in particular \citep{2009MNRAS.400.1019E,2012MNRAS.419..479E, 2016MNRAS.457.4296W}. 

Different stellar population synthesis models may incorporate different input physics and prescriptions for processes such as stellar winds. Extant studies have compared and contrasted such models when used to fit the ultraviolet-optical-near-infrared stellar emission of galaxies, identifying the strengths and limitations of different model grids \citep{2016MNRAS.457.4296W}. Dependencies on highly uncertain input assumptions within such models (e.g. star formation history and dust attenuation law) impact the accuracy of derived physical properties of galaxies \citep{2020ApJ...904...33L}. Previous work \citep[e.g.][]{2015ApJ...808..101M} demonstrates that different SPS fitting procedures can derive different values for the derived properties of galaxies. However, the effects of combining specific SPS models with different dust emission models when fitting a full spectral energy distribution, spanning from the ultraviolet (UV) to the mid-infrared (IR), have been largely neglected.

Optical galaxy SEDs contain information about the properties and evolution history of their stellar populations. However, further information is contained within other regions of the electromagnetic spectrum. Models for the IR emission component of galaxy SEDs have been developed in parallel to stellar population models. Recently, as large multiwavelength datasets (including space-based observations) have become available, IR emission models have been combined with UV, optical and near-IR stellar emission models to self-consistently predict the far-UV to the far-IR \citep[see e.g.][and references therein]{2013ARA&A..51..393C}. 

The IR emission is dominated by thermal continuum radiation from dust grains. Dust grains play an important role in galaxies, affecting the chemistry of gas in the interstellar medium (ISM), the dynamics of star formation and attenuation of short wavelength radiation from stars \citep{2011piim.book.....D}. This is absorbed by both dust and ionized gas, and is reemitted by dust in the infrared as modified blackbody emission \citep[for a review see][]{2003ARA&A..41..241D}. It has been estimated that around 30 per cent of stellar energy output is reradiated by dust in local galaxies \citep[e.g.][]{2002MNRAS.335L..41P, 2011ApJ...738...89S, 2016A&A...586A..13V}, and this fraction is higher in the dust-rich galaxies identified as infrared-luminous sources.

Young stellar populations are born within giant molecular clouds \citep[for a review on the theory of star formation see][]{2007ARA&A..45..565M}; denser regions of gas and dust found within galaxies. These birth clouds dissipate after a period of some millions of years, as radiative pressure and heating from young stars disperses the gas and dust \citep[e.g.][]{2020SSRv..216...50C,2020MNRAS.493.2872C}. Despite their short lifetimes, massive stars dominate the energy budget in most young stellar populations. Their high density birth clouds lead to higher attenuation of this hot starlight when compared to the more diffuse dust in the ISM \citep[e.g.][]{2000ApJ...539..718C}. This UV-dominated starlight is reemitted at longer wavelengths making hot birth clouds an important element of typical dust emission models.

Dust grains in the circum- and inter-stellar medium also reprocess ultraviolet stellar radiation more efficiently than optical light \citep[e.g.][]{2000ApJ...533..682C}. Since young stellar populations of ages <10~Myr dominate the UV emission from a population, dust emission is thus a tracer of young stellar populations and can be used as a measure of the star formation history \citep{1998ARA&A..36..189K, 2003ApJ...586..794B}. Since the primary heating source for dust grains is the stellar population embedded within or surrounded by the grains, the infrared emission from a population is intimately connected to the stellar heating source. This connection between stellar and dust emission must be considered to consistently predict their joint spectral energy distribution.

A range of methods have been suggested to self-consistently combine stellar and dust emission models \citep[for a review see][]{2011Ap&SS.331....1W}. A simple approach is to use an energy-balance formalism, whereby energy absorbed by dust attenuation in the UV and optical is distributed across IR dust emission components, assuming 100 per cent efficiency and simple emission properties (i.e. blackbody radiation) \citep{1999A&A...350..381D, 2008MNRAS.388.1595D}. A more sophisticated approach uses radiative-transfer calculations employing a ray-tracing method \citep{2000MNRAS.313..734E, 2007A&A...461..445S}. In any such method, assumptions must be made regarding the distribution, composition, and usually also the temperature, of dust grains as this governs the wavelength-dependence of their reemission luminosity. However, all of the calculations modelling dust emission components to date have been undertaken assuming SPS radiation fields derived from single star evolution alone - a questionable assumption given the frequency of multiplicity amongst young, massive stars. In this work, we investigate whether there is a connection between the choice of stellar population synthesis and dust emission models, and the derived galaxy properties. We explore this by fitting the photometry of luminous infrared galaxies in the local Universe as a test case. 

This paper is structured as follows. In Section \ref{models}, we present the various stellar and dust emission models used for fitting and comparison, highlighting differences between the ultraviolet budgets of stellar models available for attenuation by dust. We also discuss the \bagpipes\ SED fitting algorithm \citep{2018MNRAS.480.4379C} that is used throughout. Section \ref{valid} presents the test sample selection and fitting procedure, together with the main quantitative results of the galaxy fits. These are discussed within Section \ref{discuss}, highlighting how well different model combinations have fitted to the observational sample. In this section, we discuss the impact of the chosen models on the derived parameters, the effect of varying birth cloud age on the resulting fit, and implications for the cosmic star formation rate density history. Section \ref{lims} presents limitations to our analysis. We present a brief summary of our conclusions in Section \ref{conc}.

Throughout we report galaxy photometry, converting this to fluxes assuming a cosmological model in which $\Omega_\Lambda =0.7$, $\Omega_M =0.3$ and $H_0 = 70$\,km\,s$^{-1}$\,Mpc$^{-1}$ \citep[the default parameters for the \bagpipes\ software,][discussed in Section \ref{bagpipes}]{2018MNRAS.480.4379C}.


\section{Models} \label{models}

The analysis in Section \ref{valid} considers the combinations of three input stellar population synthesis models with three dust reemission models. In each case an energy balance formalism is used to calculate the amount of stellar light which is first attenuated and then re-emitted by the dust. Here we introduce the models considered, and describe the SED fitting methodology.


\subsection{Stellar Models}


\subsubsection{Bruzual and Charlot stellar models} \label{bc_mods}

\citet[][hereafter BC03]{2003MNRAS.344.1000B} presents a stellar model predicting the single-star-only evolution of stellar populations at ages between $1\,\times\,10^{5}$ and $2\,\times\,10^{10}$~yr, over the spectral range from 91~{\AA} to $160~\mu$m. They used the Padova 1994 stellar evolution library \citep{1993A&AS...97..851A, 1993A&AS..100..647B, 1994A&AS..104..365F, 1994A&AS..105...29F, 1996A&AS..117..113G}, which is computed at various metallicities in the range Z~=~0.0001-0.1. This library is supplemented with thermally pulsing asymptotic giant branch (AGB) evolutionary tracks \citep{1993ApJ...413..641V}, post-AGB evolutionary tracks \citep{1994ApJS...92..125V, 1983ApJ...272..708S, 1986A&A...154..125K, 1987ApJ...315L..77W}, and unevolving main-sequence star models in the mass range $0.09 \leq m < 0.6 M_\odot$ \citep{1998A&A...337..403B}.
    
To describe the individual stellar spectra of any star, they combine together the three libraries of `BaSeL' \citep{1992IAUS..149..225K, 1989A&AS...77....1B, 1991A&AS...89..335B, 1994A&AS..105..311F, 1995ApJ...445..433A, 2002RMxAC..12..150R, 2002A&A...381..524W}, `STELIB' \citep{2003A&A...402..433L}, and `Pickles' \citep{1998PASP..110..863P, 1992ApJS...82..197F}, where the former covers the whole spectral range while the latter two have higher spectral resolution but only cover specific parts of the spectral range (3200-9500~\AA~and 1205~\AA-2.5~$\mu$m, respectively). Finally, to compute the spectral evolution of stellar populations, they use the isochrone synthesis technique \citep{1991ApJ...367..126C, 1993ApJ...405..538B} with a \citet{2003PASP..115..763C} initial mass function (IMF) prescription. These models were created to help interpret spectra gathered by modern spectroscopic surveys in terms of constraints on the star formation history and metallicities of galaxies.

There is an updated 2016 version of the BC03 models (hereafter BC16)\footnote{Bruzual and Charlot models available at: \\ \url{https://www.bruzual.org/bc03/}} which is also tested. This version incorporates updated PADOVA stellar evolutionary tracks computed with the PARSEC code \citep{2012MNRAS.427..127B} for stars with initial masses up to 350~$M_{\odot}$ \citep{2015MNRAS.452.1068C} over the metallicity range Z~=~0.0001-0.040. It also introduces a new prescription for the evolution of thermally pulsing AGB stars \citep{2013MNRAS.434..488M} and the MILES library of observed optical stellar spectra (with spectral range 3525-7500~\AA) to describe the properties of stars in the Hertzsprung-Russell diagram \citep{2006MNRAS.371..703S, 2011A&A...532A..95F}. Ultimately, this modifies how the SED spectra change with stellar age, notably altering the ionizing radiation flux evolution as highlighted in the appendix of \citet{2017MNRAS.470.3532V}. As these authors report, the amount of ionizing flux varies between the two models in part due to a change in the upper-mass limit for degenerate carbon ignition, and hence post-AGB evolution, from 5~$M_{\odot}$ in the BC03 models to 6~$M_{\odot}$ in the BC16 models. Differences in the optical and near-infrared parts of the spectra arise primarily from differences in the prescriptions for post-main sequence stellar evolution \citep{2012MNRAS.427..127B, 2013MNRAS.434..488M}.

We account for nebular emission for the BC03 and BC16 models by using the pre-installed nebular grid in \bagpipes\ \citep[][see section \ref{bagpipes}]{2018MNRAS.480.4379C}. The grid was generated using the BC16 models, following the methodology of \citet{2017ApJ...840...44B} and using the {\em Cloudy} photoionization code \citep{2017RMxAA..53..385F}. This assumes H\,{\sc II} regions with a spherical shell geometry and a fixed hydrogen density of 100~atoms~cm$^{-3}$. We fix the nebular ionization parameter as $\log U_\mathrm{neb}=-3.0$.

These BC16-derived nebular grids are applied to both the BC16 and the BC03 models, since these are inbuilt into the \bagpipes\ analysis routine. While this means that the nebular emission is not going to be strictly correct for the BC03 models, the main difference is a slight over-prediction in the strength of emission due to extra blue flux in the young stars of the BC16 model. Since the BC03 models are only considered to allow consistency and comparison with previous work, this will not affect the main results, which focus on comparisons between BC16 and BPASS models.


\subsubsection{BPASS stellar models} \label{sec:bpass}

In contrast to Bruzual and Charlot models, the Binary Population And Spectral Synthesis (BPASS) stellar evolution and synthesis models \citep{2017PASA...34...58E, 2018MNRAS.479...75S} include binary evolution in addition to the single star evolutionary pathways. These were built to explore the effects of binaries on supernovae progenitors \citep{2013MNRAS.436..774E, 2016MNRAS.461L.117E} and the observed spectra from young stellar populations, but also to provide a framework which could allow for the analysis of the integrated light from both distant and nearby stellar populations \citep{2008MNRAS.384.1109E, 2009MNRAS.400.1019E,2012MNRAS.419..479E}. These models have gone on to have a wide range of applications, including exploring the rate of compact binary mergers and their role as gravitational wave progenitors \citep[e.g. ][]{2016MNRAS.462.3302E}.

The BPASS version 2 stellar evolution models \citep{2017PASA...34...58E} are derived from a heavily modified descendent of the single star evolution prescription from \citet{1964ApJ...139..306H} used in the Cambridge STARS code and its later iterations \citep{1971MNRAS.151..351E, 2008MNRAS.384.1109E}. The binary models are full, detailed stellar evolution models, in which the interior structure of the star is modelled, but which allow for mass-loss or gain through binary interactions. The evolution of the binary separation and orbital angular momentum is also tracked. 

These are combined in a population synthesis, where we use BPASS v2.2 \citep{2018MNRAS.479...75S}. The distribution of stars is defined by an IMF based on \citet{1993MNRAS.262..545K} with a power-law slope from 0.1 to 0.5~M$_{\odot}$ of -1.30 which increases to -2.35 above this to a maximum stellar mass of 300~M$_{\odot}$. Synthetic spectra for the stars are drawn from the Kurucz models of \citet{2014ApJ...780...33C}. These are supplemented with Wolf-Rayet stellar atmosphere models \citep{2003A&A...410..993H, 2015A&A...577A..13S} and O star models \citep{1998ASPC..131..258P, 2002MNRAS.337.1309S}. Synthetic populations are generated as simple stellar populations (i.e. single aged bursts) at ages of 1~Myr to 100~Gyr in increments of log(age/yr)~=~0.1 over the spectral range from 1 to 100,000~\AA, in 1~\AA~bins. These models are produced at 13 metallicities over the range Z~=~$10^{-4}$~-~0.04.

Nebular emission for the BPASS models are again calculated using the Cloudy radiative transfer code \citep{2017RMxAA..53..385F}. This illuminates a nebular cloud with the BPASS spectra. The cloud is assumed to have electron density of 200~atoms~cm$^{-3}$ (log($n_{H}$/cm$^{-3}$)~=~2.3), a fixed nebular ionization parameter and a spherical geometry. A fixed $\log U_\mathrm{neb}=-3.0$ was used in the fitting here. For the youngest stellar populations, previous work has suggested that a $\log U_\mathrm{neb}=-1.5$ or even $-1.0$ may be appropriate when considering the hard ionization spectra of the youngest stellar populations with BPASS \citep{2016MNRAS.456..485S, 2018MNRAS.477..904X}. We did not vary $U_\mathrm{neb}$ here since it would have added an additional, poorly-constrained free parameter. The nebular emission is then combined with the original spectra to form a combined stellar and nebular SED model which is inserted into the \bagpipes\ fitting algorithm.


\subsection{Attenuation of Stellar Models}

\begin{figure*}
    \includegraphics[width=\textwidth]{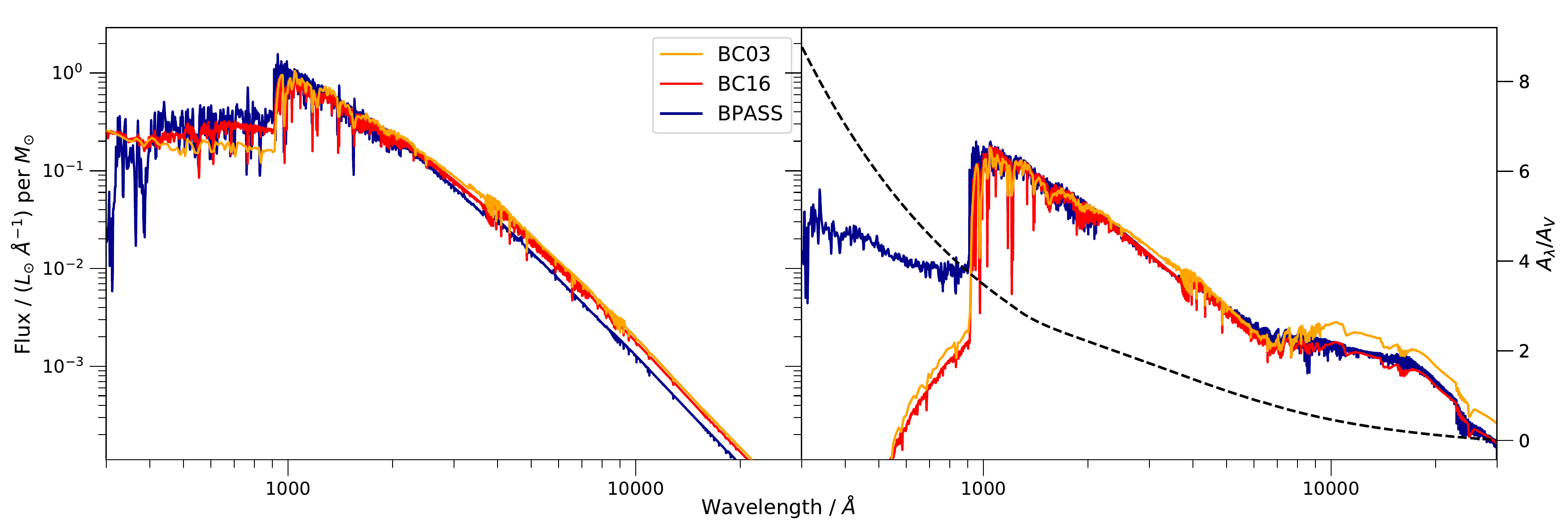}
    \caption{Comparison of the predicted SEDs from three different stellar population synthesis models: BC03 (orange), BC16 (red), and BPASS (blue). The left panel shows the SED of a 3 Myr old population while the right panel shows that of a 10 Myr old population. Due to binary inclusion in the BPASS models, they predict a lot more flux shortwards of 912~\AA\ in the 10 Myr population compared to the BC03 and BC16 models. Included for reference in the right panel as a black dashed line, with a scale to the right, is the wavelength-dependence of dust extinction \citep{2000ApJ...533..682C}.}.
    \label{fig:stel_spec_comp_wide}
\end{figure*}

Unattenuated stellar light curves for all models are plotted in Figure \ref{fig:stel_spec_comp_wide} for stellar populations at log(age/years)=6.5 and 7.0 (3.2 and 10~Myr) in the left and right panels, respectively. At 3~Myr a stellar population has a similar predicted spectrum in all the SPS models, but the same is not true for a 10~Myr-old population. Instead, the BPASS models predict a lot more flux shortwards of 912~\AA~(the ionization edge of hydrogen) due to the inclusion of binary interactions which prolongs the lifetime of O-type stars producing this ionising flux. Binary interactions therefore increase the amount of flux which can then be attenuated by dust, increasing the energy available to heat dust grains and power infrared emission.

To incorporate the attenuation of starlight by dust, the \citet{2000ApJ...533..682C} extinction law for local star-forming galaxies is applied. The dust attenuation curve takes the overall functional form of
\begin{equation}
    F_{i}(\lambda) = F_{o}(\lambda) * 10^{0.4 \frac{A_{V} k_{\lambda}}{R_{V}}},
    \label{star_attenuation}
\end{equation}
where $F_{i}(\lambda)$ and $F_{o}(\lambda)$ are the intrinsic and observed stellar flux densities, $A_{V}$ is the attenuation in the V band ($\sim$ 5500~\AA) in magnitudes, $k_{\lambda}$ is the starburst reddening curve, and $R_{V}$ is the extinction in the V band taken to be $R_{V}=4.05$. The functional form of $k_{\lambda}$ is plotted in the top panel of Figure \ref{fig:stel_spec_comp_wide}, in order to indicate the spectral regions most heavily absorbed.

\begin{figure*}
	\includegraphics[width=\textwidth]{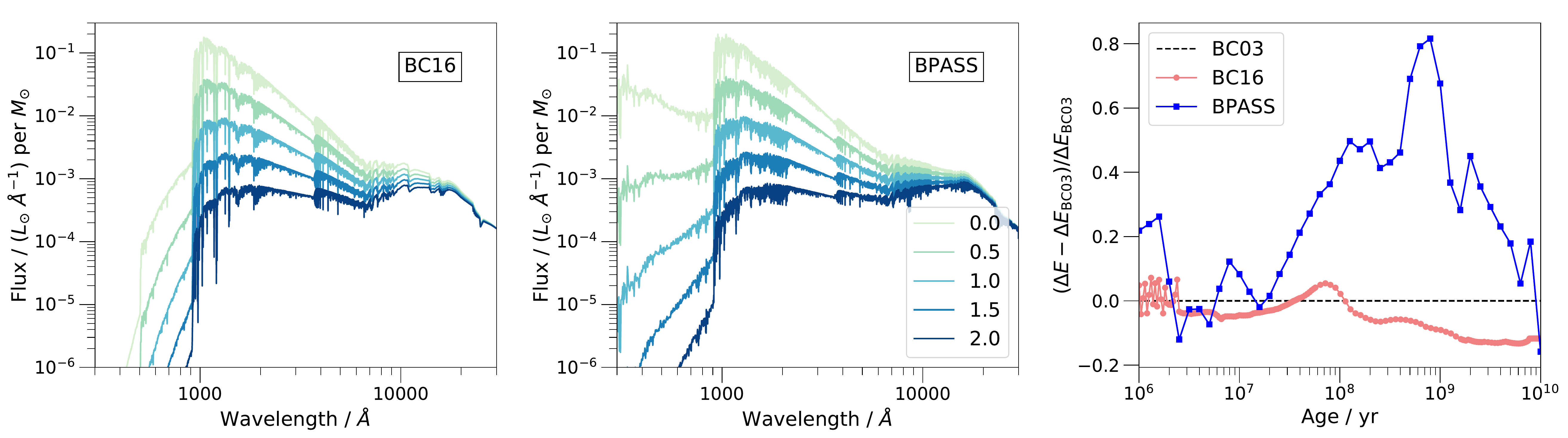}
    \caption{(Left and Centre) A 10~Myr stellar population attenuated by the \citet{2000ApJ...533..682C} extinction law at different levels of extinction for the BC16 (left) and BPASS (centre) models. The amount of extinction increases from the top, light green curve where $A_V$=0.0 (unattenuated curve) to the bottom, dark blue curve where $A_V$=2.0 in steps of 0.5. (Right) The amount of energy attenuated and transferred to the dust (${\Delta}E$) for infrared emission at different stellar population ages for an extinction level of $A_V$=1.0, shown as the fractional difference to the amount of energy transferred in the BC03 model (${\Delta}E_\mathrm{BC03}$). Plotted is the energy when considering BC03 (dashed black), BC16 (solid pink with circles) and BPASS (solid blue with squares) stellar models. Symbols indicate the time bins available in each model. Since the BC16 models contain a more finely sampled age grid, the curve varies more smoothly than the BPASS curve.}
    \label{fig:extinction_comp}
\end{figure*}

Attenuated SEDs of a 10~Myr stellar population are shown in the first two panels of Figure~\ref{fig:extinction_comp}, with the BC16 and BPASS models shown in the left and centre plots, respectively, attenuated at different levels of extinction between $A_V$=0.0-2.0 magnitudes. Attenuation is stronger at shorter wavelengths, allowing for more energy to be transferred to the dust. Thus, having young, UV emitting stars around for longer, as in the BPASS models, increases the amount of energy absorbed by dust. This is illustrated in the final panel of Figure \ref{fig:extinction_comp}, showing the amount of energy reprocessed by dust for an extinction of $A_V$=1.0 mag at different stellar ages, given as the fractional difference to the BC03 model. The BPASS models increase the energy absorbed and reemitted by dust at the majority of ages. The BC16 and BC03 models differ due to the aforementioned changes in the stellar evolution and atmosphere prescriptions. For example, the bump in the BC16 model just before 100~Myr arises from the change in the upper-mass limit for post-AGB evolution, changing the turnoff age from $1\times10^8$~yr in the BC03 models to $6.7\times10^7$~yr in the BC16 models \citep{2017MNRAS.470.3532V}. 

\subsection{Dust Emission Models} \label{dust_mods}


\subsubsection{da Cunha 2008 empirical model}

An empirical, physically motivated prescription for the thermal emission from radiation-heated dust grains was presented in \citet[][hereafter dC08]{2008MNRAS.388.1595D}. Using the BC03 models and an attenuation law following the sightline-averaged model of \citet{2000ApJ...539..718C}, they determine the amount of energy transferred into dust emission using an energy balance formalism. This assumes that all of the energy in photons absorbed by dust is reemitted with 100 per cent efficiency, but with a dependence on the dust temperature. They then split the energy across multiple emission components. These are a line-emission spectrum generated by transitions in poly-aromatic hydrocarbons (PAHs), and three thermal continuum emission components: one in the mid-infrared generated by a hot dust component, and two at longer wavelengths originating from warm grain and cold grain components. PAH emission was incorporated using an empirical spectral template derived from a local starburst galaxy (M17), while the other components are modelled as greybody emission, given by Equation 10 in dC08. In a greybody, the thermal blackbody, described by the Planck function, is modified by a wavelength-dependent dust mass absorption coefficient, $\kappa_{\lambda}$, usually approximated as $\kappa_{\lambda} \propto \lambda^{-\beta}$, where $\beta$ is the dust emissivity index. 

The relative contribution of these components is determined by the properties of the environment. In the dC08 model, the youngest stars (ages $< 10$\,Myr) are considered to be embedded in a birth cloud, which lacks the coolest emission component. However, all stars are also affected by interstellar medium (ISM) dust, from which emission representing all four components are present. These two populations are combined with the attenuated stellar emission to create an SED which consistently models emission from the UV through to the far-infrared. 

In this work, we modify the \citet{2008MNRAS.388.1595D} dust emission prescription to generate an empirical model for dust emission, given a total energy budget. The mid-infrared emission is modelled using the sum of two equally weighted greybodies with temperatures 130 and 250~K, and $\beta=1$. The warm and cold grains are described as greybody models with varying temperatures in the range 25-70~K and 12-35~K, with $\beta=1.5$ and $\beta=2$ respectively. The chosen values of $\beta$ match those of dC08, but we consider a larger temperature range. We also only fit for one warm grain temperature for both the birth cloud and ISM dust component, to reduce the number of parameters.

Instead of using the dC08 PAH emission template, we use an updated template spectrum from \citet{2021MNRAS.503.2598B}. They use archival infrared data from Herschel \citep{2010A&A...518L...1P} and Spitzer \citep{2004ApJS..154...18H} of 100 local (z~<~0.3) AGN host-galaxies selected from the 105-month Swift-BAT X-ray survey \citep{2018ApJS..235....4O} to inform a new set of infrared emission templates. In order to extract the intrinsic AGN IR emission, they initially build star-forming galaxy emission templates from 55 galaxies pre-selected using the equivalent width of the PAH emission feature at 6.2~$\mu$m in Spitzer data and WISE W1 and W2 colours. They use this galaxy selection to build eight templates for star-forming galaxies, one of which represents galaxy PAH emission and this is the template we use in our model for PAH emission.

When fitting this empirical model to observational data, all but one of the relative contributions are fitted for in each of the birth cloud and ISM components, while the last contribution is taken to bring the total energy budget to unity. We take the mid-IR emission to be the dependant variable, leaving the model with seven free parameters: the warm grain temperature, T$_W$; cold grain temperature, T$_C$; PAH contribution in birth cloud and ISM, $\zeta_{PAH}^{BC}$ and $\zeta_{PAH}^{ISM}$; warm grain contribution in birth cloud and ISM, $\zeta_{W}^{BC}$ and $\zeta_{W}^{ISM}$; and the cold grain contribution in the ISM, $\zeta_{C}^{ISM}$.


\subsubsection{Draine and Li 2007 model}

In addition to the empirical dust emission model, two grids of spectral energy templates from the literature are included, one of which is the established dust model by \citet[][hereafter DL07]{2007ApJ...657..810D}. In contrast to the empirical model, these do not assume independent dust emission components of different temperature. Instead a dust cloud is irradiated by a stellar population, assuming a distribution in dust composition, geometry and ionization parameter. They assume a dust composition mixture of carbonaceous and amorphous silicate grains, with embedded PAH material which is consistent with spectroscopic observations of PAH emission in nearby galaxies \citep{2007ApJ...656..770S}. The size distribution of the grains is consistent with the observed wavelength-dependent extinction in the local Milky Way \citep{2001ApJ...548..296W} while the abundance of PAHs and the strength of the radiation field are allowed to vary. The starlight intensity is defined with the dimensionless scaling factor, $U$, which is given by
\begin{equation}
    u_{\nu} = Uu_{\nu}^\mathrm{ref},
    \label{dl07_star_intense}
\end{equation}
where $u_{\nu}$ is the energy density per unit frequency of the starlight radiation heating the grains, while $u_{\nu}^\mathrm{ref}$ is the interstellar radiation field estimated by \citet{1983A&A...128..212M} for the solar neighbourhood. 

The bulk of the dust will be heated by a diffuse radiation field generated by many stars, but there will be some dust which will be located closer to luminous stars and thus receive a more intense radiation field. To account for this, they use a power-law distribution in intensity between $U_{min}<U<U_{max}$ for a subset of the dust, while the rest of the dust receives radiation intensity $U_{min}$. This effectively reproduces the properties of the dC08 model birth cloud, which is exposed to more intense radiation and more strongly heated. When comparing to the SED of galaxies in the SINGS field \citep{2003PASP..115..928K}, \citet{2007ApJ...663..866D} find that the data is reproduced satisfactorily with a power-law slope in $U$ of~-2 and $U_{max} = 10^6$. These were adopted as canonical parameters. Thus, their model's dust emission spectrum shape is determined by three free parameters: the PAH mass fraction, q$_{PAH}$; the lower cutoff of the starlight intensity distribution, $U_{min}$; and the fraction of the dust strongly heated by starlight with intensity $U~>~U_{min}$, $\gamma$.


\subsubsection{Draine 2021 model}

A newly released, updated set of templates provided by \citet[][hereafter D20]{2021ApJ...917....3D} are also considered. These are conceptually similar to the DL07 models, and contain similar assumptions. The overall dust composition is modified to be dominated by a mixture of amorphous silicate, other metal oxides, and hydrocarbons, a material defined by \citet{2021ApJ...909...94D}. A simple power-law is used to capture the size distribution of the grains. There is also a PAH population and an additional carbonaceous material added to fully describe the total dust emission, where the PAH size distribution is allowed to vary. Since interstellar PAHs are present in a range of charge states \citep[e.g.][]{2009ApJ...697..311B, 2017ApJ...836..198P}, model grids were generated at low, standard and high ionisation states. 

D20 consider several different descriptions for the starlight irradiating the dust. These include BC03 models and BPASS models, at various population ages from 3~Myr to 1~Gyr. D20 also consider a solar neighbourhood spectrum representative of the typical radiation field in the diffuse ISM. This is used as a reference for the heating rate parameter; a dimensionless intensity parameter of a radiation field, $U$, given as
\begin{equation}
    U = \gamma_{*}\frac{u_{*}}{u'_\mathrm{ref}},
    \label{d20_star_intense}
\end{equation}
where $\gamma_{*}$ is the spectrum-averaged dust grain absorption cross section relative to the same quantity for the reference starlight spectrum modified from \citet{1983A&A...128..212M}, $u'_\mathrm{ref}$ is the integrated energy density for this spectrum, and $u_{*}$ is the energy density of the incident radiation field. This leaves four free parameters: the incident starlight radiation model, the heating rate parameter U, the PAH size distribution, and the PAH ionisation.

We adopt the publicly released model grids. We use the BPASS-irradiated models with BPASS stellar SEDs, while the BC03-irradiated templates are combined with the BC03 and BC16 stellar SEDs. Since the irradiation varies with stellar age, and dust emission models are not available for all ages in the stellar population synthesis, some manner of reconciling these is required. Here we use the dust models irradiated by a 3\,Myr-old stellar population to describe birth cloud dust emission, while models irradiated by 1~Gyr-old stars were taken as typical of the rest of the dust emission from the ISM. Since the birth cloud and ISM dust have different heating sources and intensities, we allow them to have different heating rate parameters, $U$, while the PAH size distribution and ionisation are constrained to be the same to reduce the number of free parameters. Therefore, this model has four free parameters: two heating rate parameters, $U_{BC}$ and $U_{ISM}$ (one for birth cloud, one for ISM dust); the PAH size distribution; and the PAH ionisation.


\subsection{SED Fitting Algorithm} \label{bagpipes}

Stellar and dust models were combined into composite stellar populations by utilising the Bayesian Analysis of Galaxies for Physical Inference and Parameter EStimation program \citep[\bagpipes,][]{2018MNRAS.480.4379C}. This program contains Bayesian spectral fitting code combining the effects of stellar models, nebular emission, dust attenuation and dust emission to model the integrated light from a galaxy from the far-UV to the microwave regimes. 

\bagpipes\ accepts pre-defined SPS models in the form of a grid of simple stellar-population models at a range of different ages and metallicities. From this, for a given total mass and metallicity, it creates complex populations by the application of a star formation history (SFH), which can be constructed out of one or more analytic functions. Nebular emission is included with the stellar component. An energy balance formalism can then be applied to the current emission model to account for dust attenuation and to calculate the energy re-emitted by dust. Combining these two together, \bagpipes\ gives the total (stellar + nebular + dust emission) SED for the specified input parameters. When fitting to observational data, a prior probability distribution must be specified for any parameter being varied. The \bagpipes\ output is a posterior probability distribution for each fitted parameter, together with the Bayesian Likelihood for their combination \citep[see][]{2018MNRAS.480.4379C}.

A likelihood function, $\mathcal{L}$, describes the probability of obtaining some observational data as a function of the parameters of the chosen statistical model; it gives how well the parameters explain the data. Constructed assuming uncertainties are Gaussian and independent, the likelihood function used in \bagpipes\ is given by \citep[e.g.][]{2010arXiv1008.4686H}
\begin{equation}
    \mathrm{ln}(\mathcal{L}) = -0.5\sum_{i}\mathrm{ln}(2\pi\sigma_{i}^{2}) - 0.5\sum_{i}\frac{(f_{i}-f_{i}^{\mathcal{H}}(\Theta))^{2}}{\sigma_{i}^{2}},
\end{equation}
where $f_{i}^{\mathcal{H}}(\Theta)$ is the \bagpipes\ predicted model corresponding to observed fluxes $f_{i}$ with associated uncertainties $\sigma_{i}$. In order to get the most probable parameterisation of the statistical model, the likelihood is maximised. This defines the resulting parameter space chosen by \bagpipes.

To compare models with different free parameters and constructions, we used an estimate of goodness of fit given by the Bayesian Information Criterion (BIC). This is defined as $\mathrm{BIC} = kln(n) - 2ln(\mathcal{\widehat{L}})$, where $k$ is the number of free parameters in the model, $n$ is the number of observations, and $\mathcal{\widehat{L}}$ is the maximum value of the likelihood function returned by \bagpipes. The BIC is minimised by the best fitting model to a given set of data, and penalises models with additional free parameters. We use this to discuss comparisons between dust models with different parameterisations. We note that all the trends and interpretations reported here are also seen when the Bayesian evidence is instead used as a figure of merit.


\section{Application to Infrared-Luminous Galaxy Sample} \label{valid}


\subsection{Test Sample Data}

We validate stellar and dust emission model combinations by fitting the same set of data with the models described above, allowing a direct comparison of the quality of fit on a single uniform data set. For this purpose we consider a sample of galaxies with data extending from the UV to far-IR. These are selected from Cosmic Evolution Survey (COSMOS) observations \citep{2007ApJS..172....1S} which have been compiled into the COSMOS2015 catalogue \citep{2016ApJS..224...24L}. This survey is selected due to its wide spectral coverage which includes GALEX UV data; CFHT/Megacam, and Subaru/SuprimeCam optical data; and CFHT/WIRCam, Spitzer/IRAC, Spitzer/MIPS, Herschel/PACS and Herschel/SPIRE infrared data. This gave a spectral coverage from 1549~\AA~ to 500~$\mu$m in the observed frame. We do not include Y, J, H and Ks UltraVISTA observations as these did not cover the whole of the COSMOS field.

The COSMOS2015 catalogue is selected as it contains an unbiased sample of galaxies out to redshifts of 6. This includes a wide range of galaxy types, and characterises the kind of analysis likely to be performed over a large range of redshift in the future. We note that other galaxy samples exist which explore the thermal far-infrared of local galaxies in exquisite detail, characterising their dust properties on a morphologically-resolved basis and at high signal-to-noise. We do not use these as a validation sample at this stage, since such high quality data would be atypical of the majority of galaxy surveys for which SED fitting is routinely implemented. While COSMOS is itself a high quality data set, it consists of fewer data points and lower signal to noise for any given galaxy SED than is seen in very local samples such as DustPedia \citep{2017PASP..129d4102D}.

In addition to the photometric data, the COSMOS2015 catalogue includes value-added information such as galaxy stellar mass and star formation rate, based on a template-fitting approach carried out in the optical and using the BC03 stellar population synthesis library, together with a fitted star formation history. The catalogue also includes photometric redshift estimates. These were calibrated against the zCOSMOS survey as described in \citet{2016ApJS..224...24L}, which assesses both the typical scatter of photometric versus spectroscopic redshifts and the catastrophic failure rate as under 1\,per\,cent.

To gather validation galaxies from the catalogue, selection cuts were made. Galaxies which were flagged as having missing flux due to  being close to an image boundary or data collection being incomplete in the SuprimeCam V-band data were removed. We restrict the sample to galaxies with a good spectral coverage in the infrared. We require significant 100, 160, 250 and 350~$\mu m$ Herschel PACS and SPIRE filter detections, with a minimum signal-to-noise ratio of 3 in each filter. The 500~$\mu m$ SPIRE data was not included in this criteria as few galaxies showed strong detections in this filter. We also require full coverage in the UV, optical and infrared bands, a photometric redshift estimate and a derived mass estimate from BC03 template fitting, in order to permit galaxy subsamples to be established as a function of mass and redshift, minimising evolutionary effects. Only local galaxies ($z<0.5$) were selected, primarily due to the relatively shallow Herschel data, but also since previous dust models have been calibrated with local galaxy samples. This selection procedure reduced the sample of COSMOS galaxies to 419.

The remaining sample was checked for signs of active galactic nuclei (AGN) since we are not modelling AGN emission. \citet{2009ApJ...696.1195T} classified sources as AGN if $ L_{0.5-10~keV} > 3 \times 10^{42}$\,ergs\,s$^{-1}$. We apply this criterion to both Chandra and XMM/Newton observations collected in the COSMOS2015 catalogue. If the total X-ray luminosity in the 0.5-10~keV region from either observations exceeded this criterion then that galaxy was classified as AGN. Only one galaxy was identified as an AGN by this criterion and therefore removed.

Thus the total sample considered in this analysis is 418 infrared-luminous galaxies at $0<z<0.5$.


\subsection{Methodology}

\begin{figure}
	\includegraphics[width=\columnwidth]{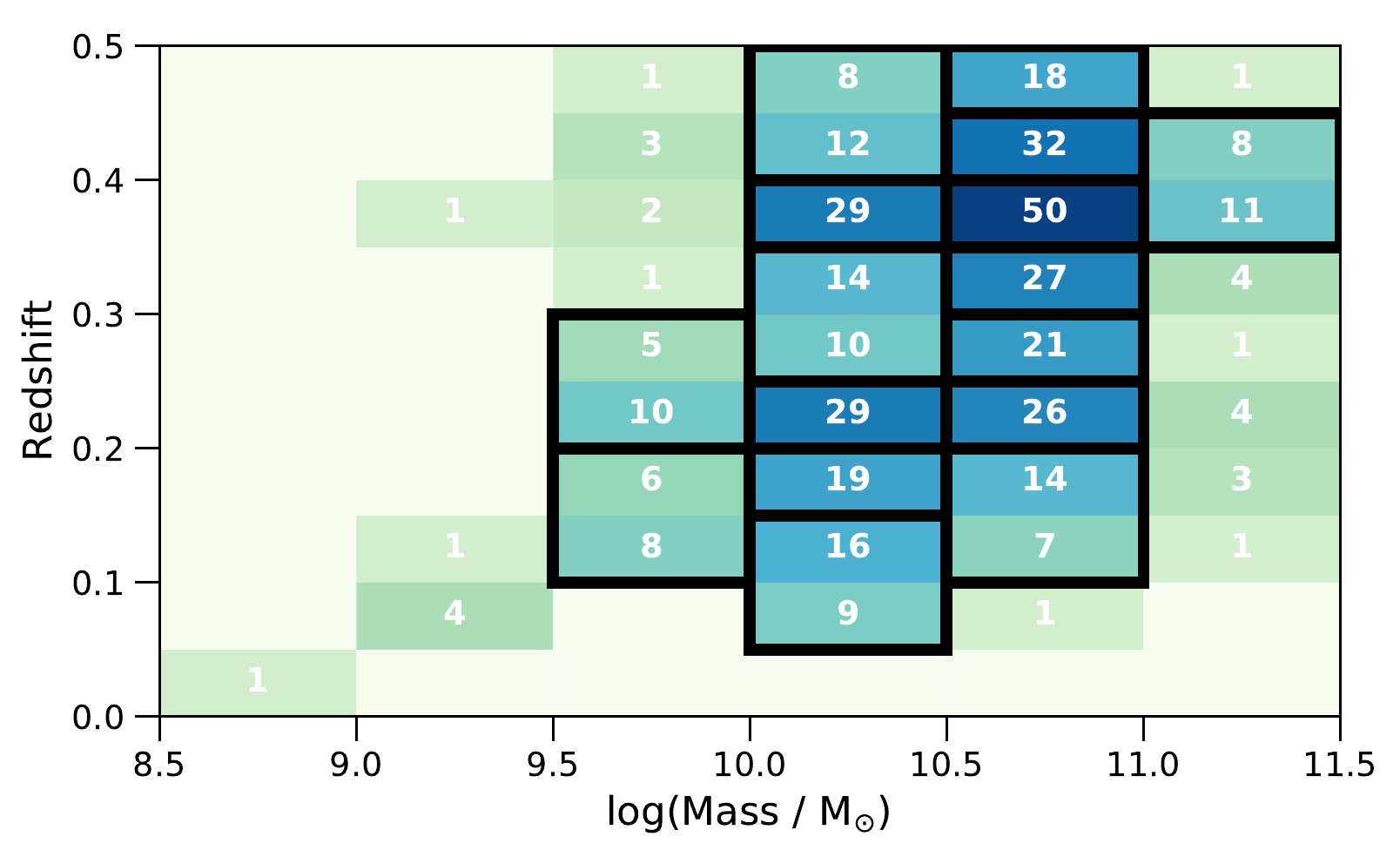}
    \caption{The mass-redshift parameter space covered by infrared-luminous galaxies in the COSMOS sample. Colour indicates a density map of the parameter space. The number of galaxies within a bin increases as the colour shifts from light green to dark blue, with the total given by the number overlaid on each bin. Black rectangles indicates subsamples which stacked into averaged spectral energy distributions and fitted with models. Where a rectangle covers two bins, it indicates that those bins were combined into one stack for fitting.}
    \label{fig:galaxy_bins}
\end{figure}

Galaxies were binned by mass and redshift. We produce stacked photometry for each bin, increasing the signal to noise (S/N) on the mean photometric data points, and also allowing the creation of a `typical' galaxy SED for that bin. This allowed modeling with a simple parametric star formation history (SFH), which averages over the typically stochastic bursty star formation in any one galaxy. Stacking also mitigates against any remaining low-luminosity AGN in the sample. Galaxies were binned by $\Delta z = 0.05$ in the range $0<z<0.5$, and further subdivided in mass, with $\Delta$(log(mass~/$M_{\odot}$))\,~$=~0.5$ in the range $10^{8.5-12.0}$\,M$_\odot$. Several of the resulting bins had low number statistics. To mitigate this, we occasionally combine adjacent pairs of redshift bins. This resulted in 16 independent galaxy stacks, which are illustrated by black rectangles in Figure~\ref{fig:galaxy_bins}; those covering two bins indicate that these were combined into one stack. A handful (29) of galaxies lie in sparsely populated regions of parameter space. These were omitted from the final fitting procedure. 

Within each bin, observational data for each galaxy (and its derived properties from previous SED fitting) are taken from the COSMOS2015 catalogue and combined with a luminosity weighting based on the B-band SuprimeCam flux. This blue optical filter weighting favours relatively young or star-forming galaxies, and thus those with energy which can be redistributed to the infrared by dust. 

A correction was made to account for Galactic extinction. \citet{2016ApJS..224...24L} calculate extinction for each object's line-of-sight using the \citet{1998ApJ...500..525S} values. We use a \citet{1999PASP..111...63F} interstellar attenuation law, for which $R_V=3.1$. Relative to other equatorial fields, the main COSMOS field was chosen as it has an exceptionally low and uniform galactic extinction \citep{2007ApJS..172....1S}, so we take the average extinction for each stacked galaxy group before applying any correction to the stacked photometry.

To test the stellar and dust model combinations discussed in Section \ref{models} against these stacked observations, composite population models were created and then fitted with \bagpipes. These are formed of two stellar populations: an old stellar population of a fitted age $>0.1$~Gyr and a less massive young stellar population with fixed age $5$~Myr. Both of these are modelled using a simple parametric, delayed-tau SFH, which defines a star formation rate, SFR $\propto te^{-t/\tau}$, where $t$ is the time elapsed between the onset of star formation and the epoch of observation and $\tau$ is a parameter describing the exponential decay timescale of the star formation rate. The age of the old stellar population was fitted for along with the mass and $\tau$ parameter of both populations, while the metallicity was fixed at Solar ($Z_{\odot}=0.02$) to reduce the number of free parameters. SED fitting is typically only weakly dependent on assumed stellar metallicity. We also select a single dust attenuation law \citep{2000ApJ...533..682C} and birth cloud attenuation multiplicative factor, $\eta$. Attenuation in the V-band, $A_{V}$, was allowed to vary. Our procedure is summarised in Table~(\ref{tab:stel_params}) which gives the fitting ranges for all the variable parameters. 

The uncertainties on photometric flux measurements are typically very small. For fitting with discrete grids of photometric models, this is problematic since it is very possible that no single model will pass through the very narrow range of measurement uncertainties. To account for overall calibration uncertainties and finite model sampling, the photometric uncertainties were increased, where required, to a maximum S/N of 20 for all filters except for the IRAC, PACS and SPIRE filters where a maximum S/N of 10 is chosen. The infrared filters have larger typical errors due to increased source confusion and hence larger systematic uncertainties.

\begin{table}
	\centering
	\caption{Parameter ranges considered when performing SED fitting with \bagpipes. If a single value is given the parameter was fixed.}
	\label{tab:stel_params}
	\begin{tabular}{lcc}
		\hline
		 & Parameter & Value/Range\\
		\hline \hline
		Old stellar & Age / Gyr & 0.1~-~13\\
		population SFH & $\tau$ / Gyr & 0.05~-~10\\
		 & log(Mass/$M_{\odot}$) & 3~-~13\\
		 & Metallicity & $Z_{\odot}=0.02$\\
		\hline
		Young stellar & Age / Gyr & 0.005\\
		population SFH & $\tau$ / Gyr & 0.05~-~10\\
		 & log(Mass/$M_{\odot}$) & 3~-~13\\
		 & Metallicity & $Z_{\odot}=0.02$\\
		\hline
		Nebular emission & $\log(U_\mathrm{neb})$ & -3\\
		\hline
		Dust attenuation & Law & \citet{2000ApJ...533..682C}\\
		 & $A_v$ / mag & 0~-~5\\
		 & $\eta$ & 2\\
		\hline
		Birth Cloud & Age & Fixed\\
		\hline
	\end{tabular}
\end{table}

For any given fitting procedure, the birth cloud age was fixed. However this was varied between runs in order to evaluate how the birth cloud affected the modelling for the different stellar populations. All model combinations were fitted with maximum birth cloud ages of either 3 or 5~Myr. These timescales are representative of the likely birth cloud dispersal ages of young stellar populations \citep[see][]{2020MNRAS.493.2872C}. In the analysis to follow, we adopt 3 Myrs for the birth cloud in BC03 or BC16 models, since in these the ultraviolet flux drops dramatically in this time period, and further increases in birth cloud lifetime will have negligible impact on the dust reemission. BPASS stellar populations remain ultraviolet-luminous for longer time periods, and we fit these with a 5\,Myr birth cloud by default. This point will be discussed further in Section~\ref{bc_life}.

We fit for the free parameters of each dust model, as described in Section~\ref{dust_mods}. The fitting ranges are specified in Table~(\ref{tab:dust_params}). The DL07 and D20 models parameter ranges are limited by the publicly-available grids. The parameter range for the dC08 model has been selected to encompass the expected properties of the galaxies in the sample.

\begin{table}
	\begin{center}
	\caption{All dust emission models fitting ranges for each parameter in the model. }
	\label{tab:dust_params}
	\begin{tabular}{ccc} 
		\hline
		Model & Parameter & Range\\
		\hline \hline
		DL07 & q$_{PAH}$ / \% & 0.5~-~4.5\\
		 & U$_{min}$ & 0.1~-~25\\
		 & $\gamma$ & 0~-~0.5\\
		\hline
		dC08 & T$_{W}$ / K & 25~-~70\\
		 & T$_{C}$ / K & 12~-~35\\
		 & $\zeta^\dagger$ & 0.001~-~0.99\\
		\hline
		D20 & log(U$_{BC}$) & 0~-~7\\
		 & log(U$_{ISM}$) & 0~-~7\\
		 & PAH size & Small, Standard, Large\\
		 & PAH ionisation & Low, Standard, High\\
		 
		\hline
	\end{tabular}\\
	\end{center}
	$^\dagger$ $\zeta_{PAH}^{BC}$, $\zeta_{PAH}^{ISM}$, $\zeta_{W}^{BC}$, $\zeta_{W}^{ISM}$ and $\zeta_{C}^{ISM}$ are fit independently, but over the same parameter range in each case.
\end{table}

\begin{figure*}
	\includegraphics[width=\textwidth]{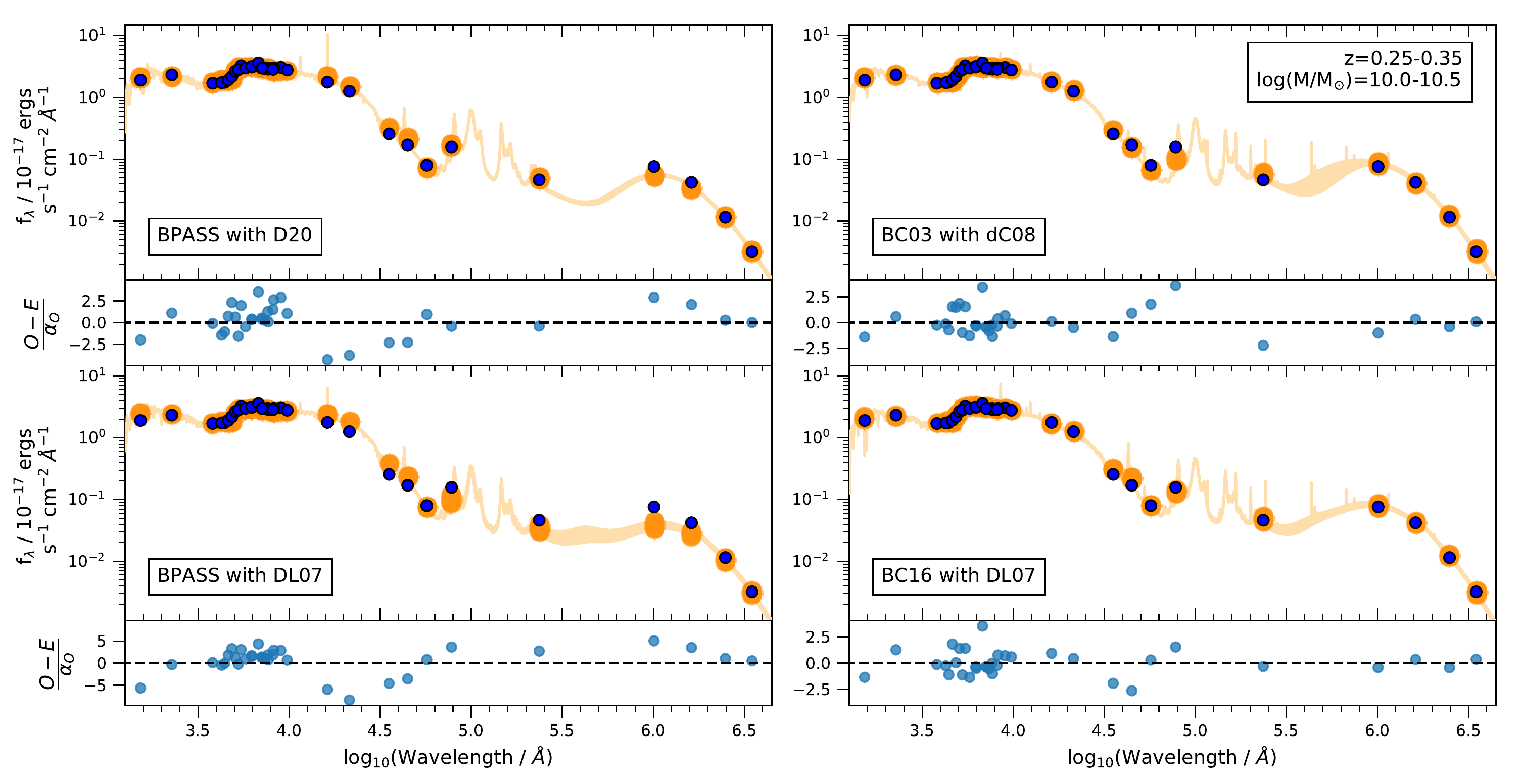}
    \caption{Model combinations fitted to stacked galaxy data from the bin at z=0.25-0.35 and log(M/$M_{\odot}$)=10.0-10.5 with a 5~Myr birth cloud age. Each fit has the best fit spectrum (orange line) plotted in the top panel with the photometric flux calculated in each filter used to constrain the model (orange dots), over-plotted with the observational data (blue dots). Errors have been included for the observational data but are too small to be seen. The bottom panel for each fit shows the normalised residuals in each photometric filter. These are calculated as the difference between the observational (O) and model (E) data, normalised by the uncertainty in the observational data ($\alpha_{\mathrm{O}}$). The model combinations shown are BPASS with D20 (top left), BPASS with DL07 (bottom left), BC03 with dC08 (top right), and BC16 with DL07 models (bottom right). Note that the models have been redshifted to the mean redshift of the bin (z\,=\,0.298). The thickness of the orange line represents the uncertainty in the model, highlighting the range of model parameters which are consistent with the observations.} 
    \label{fig:model_fits_grid_group6}
\end{figure*}


\subsection{Impact of Model Choices on Derived Parameters}\label{sec:impact}

As discussed above, we consider fits with three stellar models (BC03, BC16, BPASS) and three dust models (DL07, dC08, D20), giving nine independent combinations.

In Figure \ref{fig:model_fits_grid_group6}, we give examples of the models with the maximum likelihood when fit to galaxy observations in a bin at z=0.25-0.35 and log(M/$M_{\odot}$)=10.0-10.5, assuming a birth cloud age of 5~Myr. Fits using all nine different combinations of the 3 stellar and 3 dust models are also shown in the Appendix (Figure \ref{fig:model_fits_all_grid_group6}). The  observational data are well reproduced with the four model combinations shown in Figure \ref{fig:model_fits_grid_group6}. However, there are certain combinations which show larger residuals, such as the combination of BPASS stellar and DL07 dust models. The optical residuals are systematically offset from zero to balance the poor fit in other parts of the spectrum in this case. This is indicated on the figure by the thickness of the shaded line which gives the uncertainty and thus highlights the range of models and model parameters which are consistent with the observational dataset. Most of the uncertainty appears between the MIPS 24~$\mu m$ and PACS 100~$\mu m$ filters due to a gap in the photometric coverage. This uncertainty is associated with the dust temperature parameters (see Section \ref{dust_temp}), which for the DL07 and D20 models is defined by the incident starlight radiation intensity and fraction of strongly heated dust. Similar best-fitting spectra are found for all other mass-redshift subsamples or when using a 3~Myr birth cloud dispersal age. In the Appendix we show an example of a corner-plot (Figure \ref{fig:model_corner_plot}), illustrating this parameter degeneracy.

\begin{figure}
	\includegraphics[width=\columnwidth]{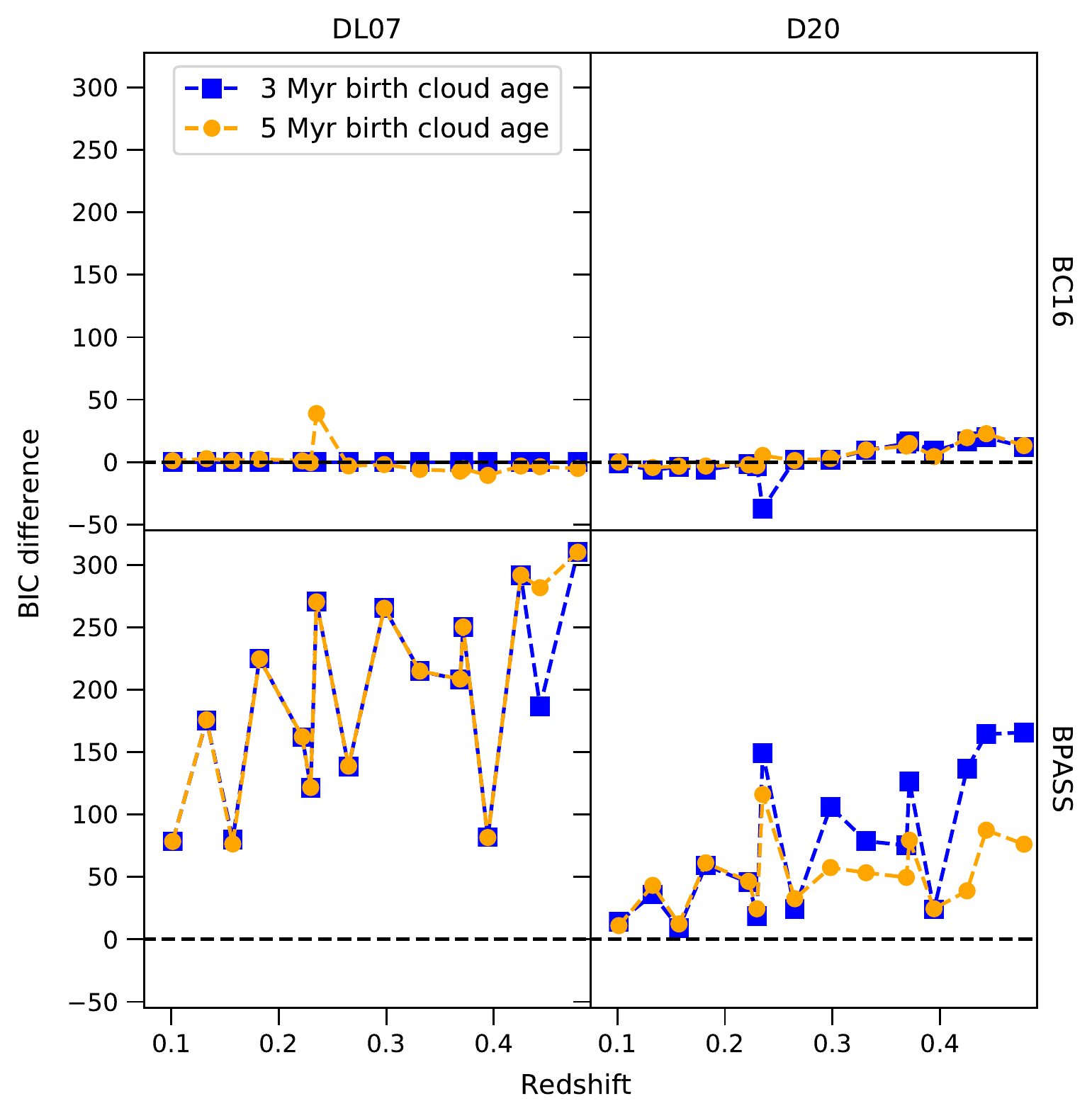}
    \caption{BIC values for a selected combination of models, calculated as the difference to the BC16 and DL07 model combination with a 3~Myr birth cloud maximum age. The top row is for BC16 stellar models while the bottom is for the BPASS models, and these are combined with DL07 dust emission models in the first column and D20 dust models in the second column. The blue squares show the resulting BICs for a 3~Myr birth cloud age while the orange circles are for a 5~Myr birth cloud age. Lower BIC values and hence lower BIC differences indicate that is the preferred model.}
    \label{fig:red_likes}
\end{figure}

We compare the BIC values of the best fits in Figure~\ref{fig:red_likes}. These are calculated as the difference to the BC16 and DL07 model combination with a 3~Myr birth cloud age. We show the BC16 and BPASS stellar models with the DL07 and D20 dust models for both a birth cloud age of 3~Myr and 5~Myr. Each point represents a different combination of mass- and redshift-range selection and is plotted at the mean redshift of the galaxies included. Lower BIC values and hence a lower BIC difference indicates  the preferred model out of the two. This highlights the poor combination of BPASS stellar and DL07 dust models, this time over all galaxy sub-selection bins fit. This combination has the highest BIC values out of all the combinations evaluated. While there is a slight increase in BIC with redshift, the BPASS and D20 model combination shows a similar goodness-of-fit to the BC16 and D20 model combination. The BPASS stellar models produce a better match to the data with a 5~Myr birth cloud age rather than a 3~Myr age. A BIC comparison plot including all nine model combinations is given in Figure \ref{fig:all_likes}. The older BC03 stellar models show similar BIC trends  to the BC16 models, while the dC08 dust models produce trends similar to the D20 grid.

\begin{figure}
	\includegraphics[width=\columnwidth]{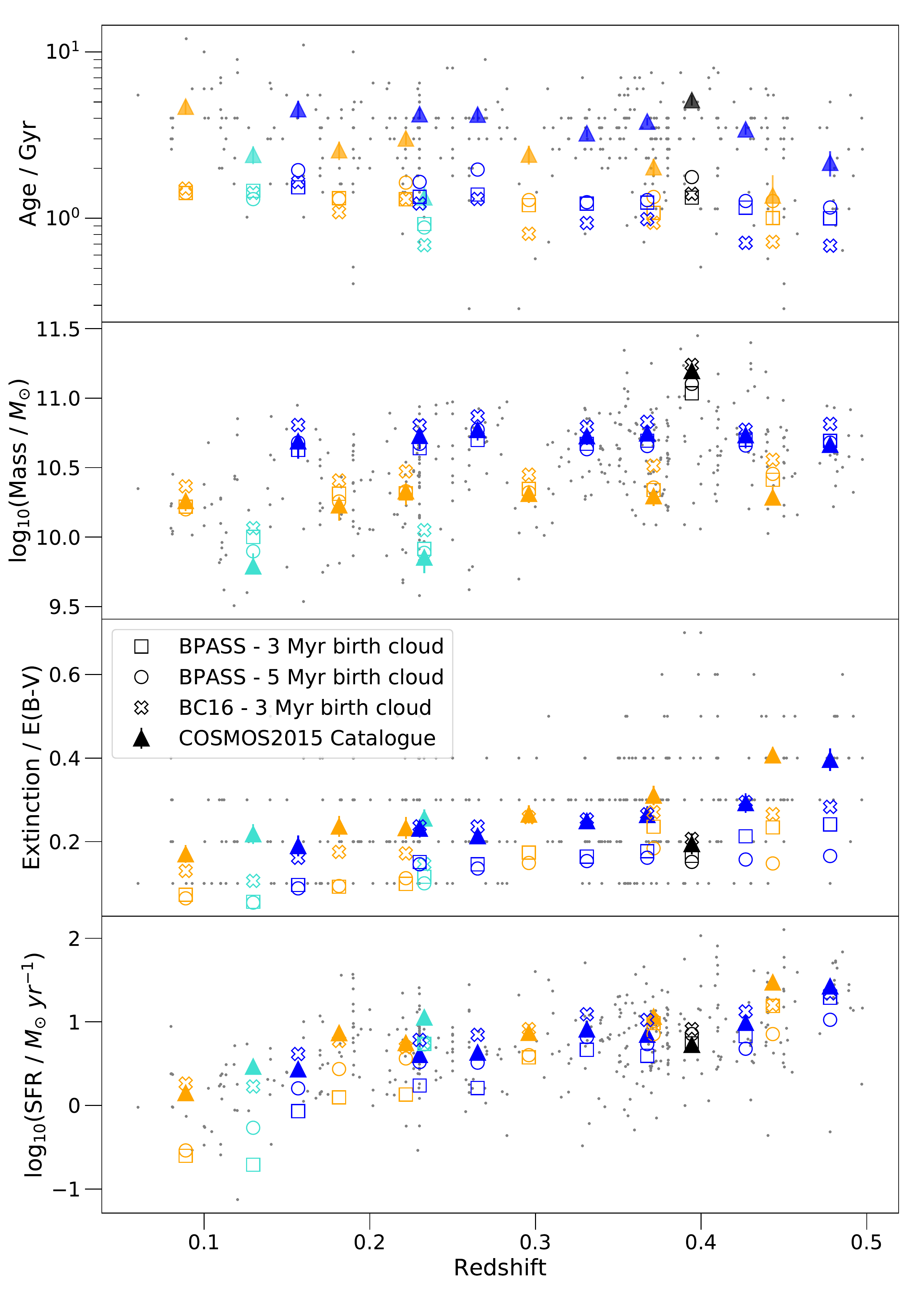}
    \caption{Resulting best-fit stellar parameters of age (top panel), mass (upper middle), dust extinction (lower middle) and SFR (bottom) from \bagpipes\ to create models which match the observations. COSMOS2015 parameters are included for individual galaxies in the sample as grey dots, and the B-band luminosity weighted average value for each sub-sample are plotted as filled triangles. The open symbols are the best-fit values from \bagpipes\ using the D20 dust model, with BPASS stellar models being represented by the square and circle for a 3 and 5~Myr birth cloud age respectively, while the cross represents the BC16 combination for a 3~Myr birth cloud. The colours represent the different mass bins for clarity, where the log(M/$M_{\odot}$) = 9.5-10.0, 10.0-10.5, 10.5-11.0, and 11.0-11.5 bins are shown in light blue, orange, dark blue, and black, respectively. Errors are included only on the COSMOS2015 values for clarity.}
    \label{fig:stel_param_comp}
\end{figure}

We compare the best-fit stellar parameter space recovered when using three of the better fitting model combinations in Figure~\ref{fig:stel_param_comp}. For comparison we also show the best-fit values from the COSMOS2015 catalogue, where physical properties were derived using synthetic spectra generated from BC03 models using a \citet{2003PASP..115..763C} IMF. Two metallicities (Solar and half-Solar) with two different attenuation curves, one using the \cite{2000ApJ...533..682C} extinction law and the other being a curve with a slope $\lambda^{0.9}$, were considered. The COSMOS fitting assumed a SFH of form $SFH=\tau^{-2}te^{-t/\tau}$, as described in \citet{2015A&A...579A...2I}. 

Figure~\ref{fig:stel_param_comp} shows the COSMOS2015 catalogue values for the individual galaxies which satisfy our selection criteria (grey dots), and the mean values weighted by B-band luminosity of each galaxy in each sub-selection bin (filled triangles). These are compared to the \bagpipes\ parameter output (unfilled symbols) when using D20 dust models with BPASS for a 3~Myr (squares) and 5~Myr birth cloud ages (circle), and when combined with BC16 models for a 3~Myr age (cross). The colours represent different mass bins for clarity.

The resulting stellar parameters show dependence on the input model assumptions. The extinction estimated using BPASS models always falls lower, by an average of E(B-V)~=~0.07, than both the COSMOS2015 and BC16 value, which are generally in good agreement with one another.  Due to the harder ionising spectrum of the BPASS models, less dust reprocessing is required to achieve the same amount of dust emission. Hence lower extinctions are favoured by the best-fitting models. In general, the 5~Myr birth cloud age models have lower derived extinctions by E(B-V)~=~0.03 on average, as expected since the longer lifetime leads to more radiation being reprocessed as dust emission. The extra flux available for reprocessing in BPASS models also permits the mass of the galaxy chosen by \bagpipes\ to be slightly lower ($0.14\pm0.02$\,dex, averaged across the whole sample) than the BC16 models for all bins, as shown in the upper middle panel. Nonetheless, stellar mass is generally recovered robustly, with typically less than 0.2\,dex of uncertainty introduced by dependence on model choices. Inferred star formation rates (SFRs) show good agreement between the COSMOS2015 fitting values and those using BC16 models. SFRs from BPASS fitting are typically lower, with an average offset of $0.31\pm0.17$\,dex. While there is some variation from subsample to subsample, the lowest mass galaxies typically show larger differences between SPS model fits and those in the highest mass bins are smaller. The age of the stellar populations show the most striking differences. Stellar ages derived from \bagpipes\ fits are reduced by a factor of 2-3 compared to the COSMOS2015 value, for the majority of bins. 

We note that we have assumed a simple parametric SFH and Solar metallicity in all cases.  \cite{2016ApJS..224...24L} use a more sophisticated approach than this, which gives rise to many of the specific differences found in the derived age and SFR of galaxies. However, the goal of this work is to evaluate the impact of stellar and dust model combinations rather than derive the precise stellar population for a given galaxy. As such, we have considered stacked galaxies in order to smooth out galaxy-specific variation, permitting us to assume a simple parametric SFH and fixed metallicity. Our focus is on the extent of differences between model outputs, rather than their specific values. 

The main result presented here is that the assumed input physics of modelled SEDs can have a considerable effect on the physical parameters derived for a galaxy, or stacked spectrum, depending on whether the input physics is the stellar models, SFH or birth cloud age chosen. For example, varying the birth cloud age between 3 and 5~Myr can alter the final physical parameters, especially the derived extinction of moderate redshift galaxies where the difference in BIC values in Figure~\ref{fig:red_likes} becomes noticeable.


\section{Discussion} \label{discuss}


\subsection{Optimising the Star-Dust Relationship} \label{star_dust_rel}

We have shown in Figures~\ref{fig:model_fits_grid_group6} and \ref{fig:red_likes} that there are combinations of stellar and dust emission models which lead to poorer matches against observations. In the specific case of BPASS stellar SEDs with DL07 dust emission, there is a large difference in fitting performance. When DL07 created their model grid, they tested them by fitting to galaxies selected from the SINGS galaxy survey \citep{2003PASP..115..928K}. This is a collection of 75 nearby galaxies with comprehensive infrared imaging to capture and characterise the infrared emission from galaxies across a range of galaxy properties and star formation environments. After fitting this sample, DL07 assumed a maximum starlight radiation intensity, $U_{max}$, and the power-law slope in the intensity distribution for their models, reducing the number of free parameters and removing the requirement to model an independent, high temperature birth cloud. The bulk of the dust (i.e. ISM component) is just illuminated by a fixed $U_{min}$. 

We have shown here that the BC03 and BC16 stellar models fit well to observations when combined with these dust models. However, such simplifying assumptions may be inappropriate when considering the BPASS stellar models. The properties of a birth cloud dust component, along with the harder ionising spectra of the BPASS models is seen to have an effect in Figure~\ref{fig:red_likes}. A longer lived birth cloud allows for more of the hard ionizing flux to be captured and thus improves the fits to the observations for the BPASS and D20 model combination. However, in the BPASS with DL07 model combination, the stellar light intensity distribution from BPASS has no effect on the shape of the dust emission model - only its total energy, rather than its range of ionization potentials, is considered. Without a birth cloud in DL07 models, there is no specific component which can reprocess the extra ionizing flux in the BPASS stellar spectra. Doing so would require additionally varying the currently-fixed stellar intensity distribution - however this would still need to be related in some way to the stellar population. 

BC03 and BC16 stellar models have considerably less ionizing flux at ages of 3-5 Myrs, such that the limited flexibility in treating young starlight has little effect. On the other hand, fits with BPASS will be particularly sensitive to this prescription in this study, especially since our target galaxies are typically highly luminous, and more intensely star forming than was true of the SINGS galaxy sample. This emphasises the need to relate the dust emission to the stellar emission SED.

Figure \ref{fig:red_likes} suggests that fits obtained with BPASS typically have slightly higher BIC values than those obtained with other stellar models. This arises primarily due to large residuals in fits to the near-infrared (1.5-5\,$\mu$m) bands. In this wavelength range, the composite emission spectrum is dominated by the combination of stellar and nebular gas emission. The poor fits here may indicate that either a modified treatment of giant branch stellar atmospheres is required in BPASS, or that the fixed nebular gas prescription used in this study is inappropriate for the COSMOS galaxies, and a much higher ionization potential component should have been included. This spectral region is highly sensitive to the details of the assumed giant branch stellar wind and mass loss prescription, and also to the details of any binary interaction prescription since it is on the giant branch that most such interactions take place. However, the BIC discrepancy appears to increase with increasing redshift, and is thus associated with more intense, younger stellar populations, which suggests that the nebular ionization parameter may instead be at fault. The nebular emission was fixed with the ionization parameter value chosen to match typical local galaxies fit previously using the lower ultraviolet flux output of the BC03 and BC16 models, meaning the same effect is not observed in the performance of those models. However, BPASS models may require a larger nebular ionisation parameter when fitting (see Section \ref{stel_assump}).


\subsection{Impact on Inferred Stellar Population Uncertainties}

Figure~\ref{fig:stel_param_comp} showed how the stellar parameter space varies depending on the input stellar model, birth cloud age or SFH parameterisation. Each of these gives similar but different results for the physical parameters of a galaxy. The mass of the galaxies is typically consistent between models at the level of $\sim$0.2\,dex, since this is primarily determined by the flux normalisation of the predicted model. The BPASS models are consistent with those derived by the COSMOS team with a typical scatter of 0.05\,dex. There is a larger typical discrepancy between the COSMOS2015 derived properties and those from our fits using the BC16 models. This arises primarily due to changes in the adopted stellar atmosphere models between BC16 and BC03, together with the different wavelength coverage of the data being fit, and different fitting methodologies. The optical continuum mass-to-light ratio, which for BPASS is lower than that of BC16 at a fixed age, has a systematic effect on the resulting mass of the galaxy. This results in BPASS mass estimates falling below those using BC16 stellar templates by, on average, 0.15\,dex, when the same fitting algorithm (\bagpipes) is used. Extinction also shows a systematic offset, since the harder ionizing spectrum of BPASS models requires less extinction to produce the same total emission spectrum. This results in the BPASS 3 and 5~Myr birth cloud age models requiring extinctions lower by E(B-V)~=~$0.064\pm0.022$ and $0.087\pm0.026$, respectively, compared to the BC03 stellar models with a 3~Myr birth cloud age.

The age of the population shows a less systematic trend than the other parameters, as the input parameter space and the history of the galaxy play a stronger role in determining this property. However, all galaxy mass and redshift bins are broadly consistent with \bagpipes\ derived stellar population ages in the range 0.5-2~Gyr.

In this study, we have adopted simple parametric star formation histories in order to permit direct comparison between different input models. A more detailed, usually non-parametric SFH is often required to fully map the evolution and merger history of any given galaxy \citep{2019ApJ...873...44C, 2019ApJ...876....3L}. However, the systematic trends identified above will persist regardless of star formation history.


\subsection{Birth Cloud Lifetimes} \label{bc_life}

The survival timescale of the dusty stellar birth cloud 
is an important parameter, since an increased survival lifetime before the birth cloud dissipates results in more young, UV luminous stars experiencing increased extinction. It thus increases the amount of energy available to be reradiated by dust. From Figure~\ref{fig:red_likes}, we have noted an improvement in fits using BPASS stellar templates when using a 5~Myr birth cloud age instead of a 3~Myr age, while no such improvement is seen for the BC16 models. 

\begin{figure}
	\includegraphics[width=\columnwidth]{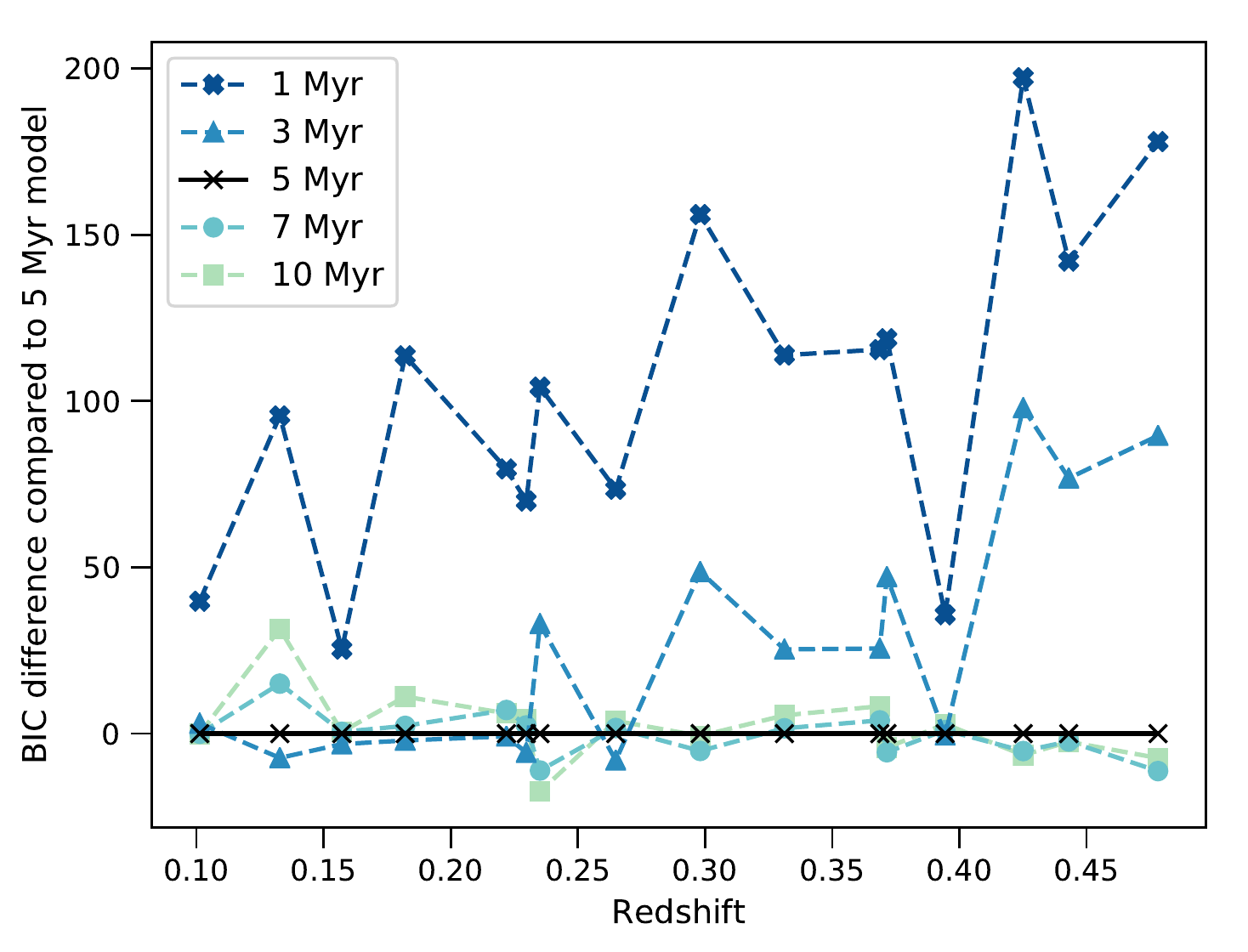}
    \caption{BIC values for all galaxy composites fitted with BPASS and D20 model combination at  assumed birth cloud ages between 1 and 10~Myr, shown as the difference to the 5~Myr model. The dashed lines are shown for clarity to indicate trends. Symbols indicate models assuming a 1 (thick crosses), 3 (triangles), 5 (thin crosses), 7 (circles) and 10~Myr (squares) birth cloud age.}
    \label{fig:like_bpass_d20}
\end{figure}

To explore this further, we ran additional fitting procedures on the BPASS and D20 model combination to explore a larger range of birth cloud lifetimes. We show the results in Figure~\ref{fig:like_bpass_d20}, which gives the BIC values for all galaxy composites, but now permitting the maximum birth cloud ages to vary between 1 and 10~Myr. The figure shows the improvement in fit quality, quantified as the difference compared to the BIC when a 5~Myr birth cloud is assumed.

Models with birth clouds that dissipate after just 1~Myr consistently show higher BICs, suggesting a worse fit. Figure~\ref{fig:like_bpass_d20} indicates that the BICs typically decrease to a minimum around 5~Myr (a few galaxy composites are best fit at 3 or 7~Myr). When the birth cloud survival age is increased beyond this, there is little variation or improvement in the fits. Hence a birth cloud, its correct lifetime and the extra extinction it causes is clearly a required component of physically accurate models.

As Figure~\ref{fig:red_likes} indicated, the sensitivity to birth cloud lifetime is particularly acute when using BPASS stellar population templates and when looking at the very young stellar populations which dominate with increasing redshift. This is unsurprising. A BC16 stellar population at 3\,Myr has already emitted the vast bulk of its ultraviolet photons, and hence birth clouds with ages less than this dominate the energy budget for dust emission. Increasing the birth cloud survival age does not substantially increase the available reprocessed energy. By contrast, BPASS populations continue to be strongly ionizing, and to have luminous ultraviolet emission to ages exceeding 10\,Myr. Thus, neglecting stars at slightly higher ages potentially leads to an underestimate of the integrated dust emission luminosity. Beyond 5\,Myr, further increase to the birth cloud lifetime results in no further improvement of the fit. Instead, this results in a large amount of reprocessed ultraviolet energy and hence excess dust emission beyond that required by the data; as a result, the galaxy SED fitting process favours older stellar populations. This reduces the available ultraviolet flux that can be reprocessed, to compensate for the more efficient reprocessing.

The stellar population at ages of 5-10\,Myrs has still {\em not} exhausted its ultraviolet energy budget, which instead declines slowly and continuous between 3 and 10\,Myrs (as shown in Figure~\ref{fig:stel_spec_comp_wide}). Thus this behaviour cannot be attributed to the stellar population. The observed trends in fitting quality may, then, indicate that the intrinsic lifetime of dense, highly extincting, dusty birth clouds is of order 5 Myrs. After this, radiative pressure and heating from the young stellar population dissipates the dense birth cloud. Previous work has investigated the lifetime of birth clouds and found similar ages of 1-5~Myr for their dissipation after massive stars formed \citep[e.g.][]{2015MNRAS.449.1106H, 2017A&A...601A.146C, 2020SSRv..216...50C}. While this is still debated and molecular clouds can have lifetimes of several tens of Myrs \citep[e.g.][]{2009ApJS..184....1K, 2012ApJ...761...37M}, the key is how quickly they disperse after the massive stars formed, as these will dominate the UV emission and hence the infrared emission from birth clouds.

In summary,  using too short a birth cloud lifetime forces an SED-fitting algorithm to increase the attenuation to match any observed far-IR fluxes, worsening the fit to the UV-NIR continuum. This is particularly problematic for SPS models in which the UV-luminous lifetime of stars is extended.


\subsection{Warm Grain Temperature Variations} \label{dust_temp}

\begin{figure}
	\includegraphics[width=\columnwidth]{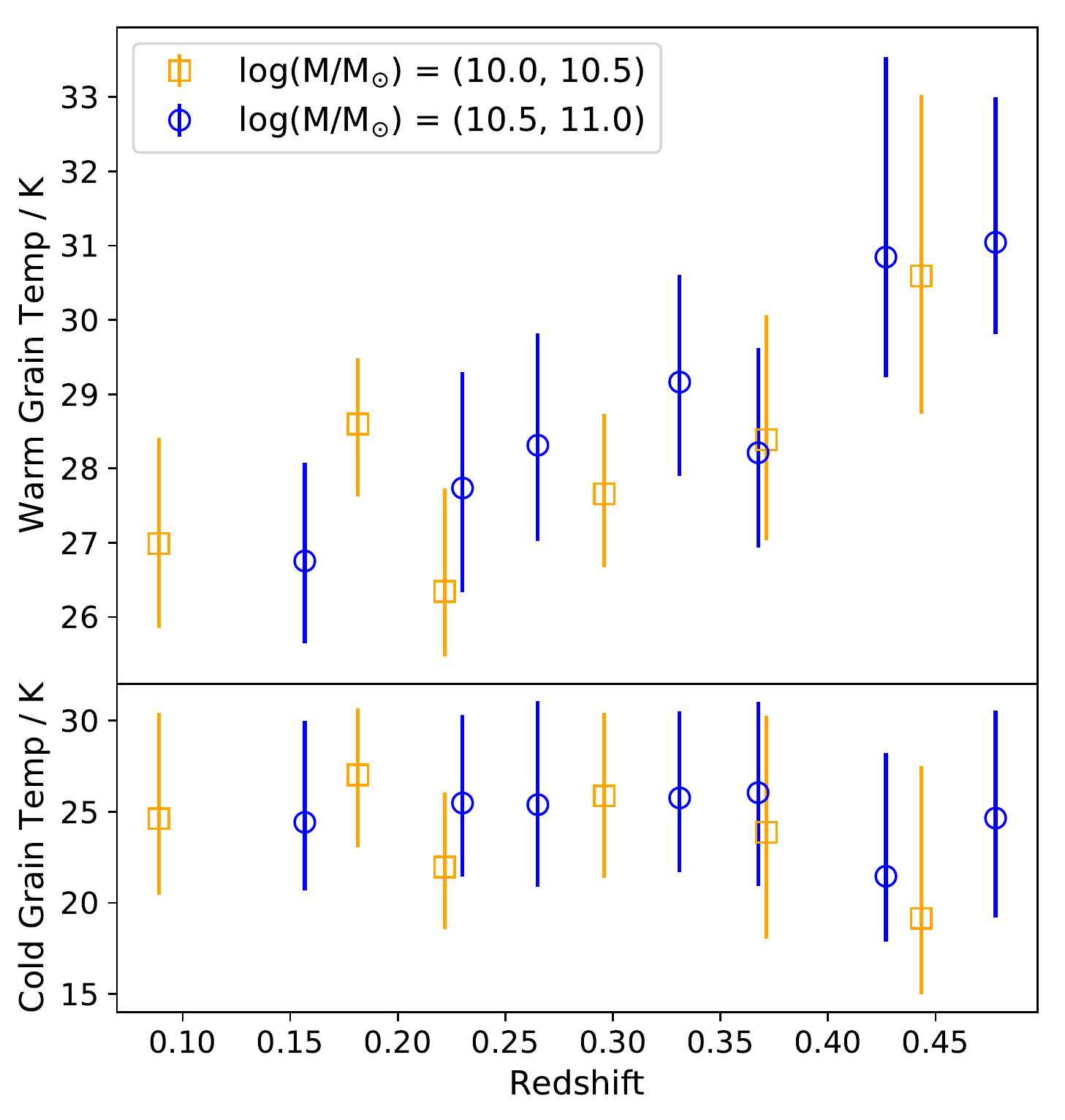}
    \caption{Best fit parameters for the dust grain temperatures of the dC08 model when fitted with BPASS stellar models with a birth cloud age of 5~Myr. The top panel is the best fit temperature for the warm grain component, while the bottom panel is the temperature for the cold grain component. The symbols represent the median value of the probability distribution, while the errors equate to the 16 and 84 percentiles of the distribution. Only the results for the log(M/$M_{\odot}$)=10.0-10.5 (orange squares) and 10.5-11.0 (blue circles) galaxy stacks are shown to remove effects relating to the mass of the galaxy samples, and the other mass ranges contain too few redshift bins to draw any conclusions. Note that the warm grain minimum fitting temperature is 26~K while the cold grain maximum temperature is 30~K.}
    \label{fig:cunha_temp_bpass}
\end{figure}

\begin{figure}
	\includegraphics[width=\columnwidth]{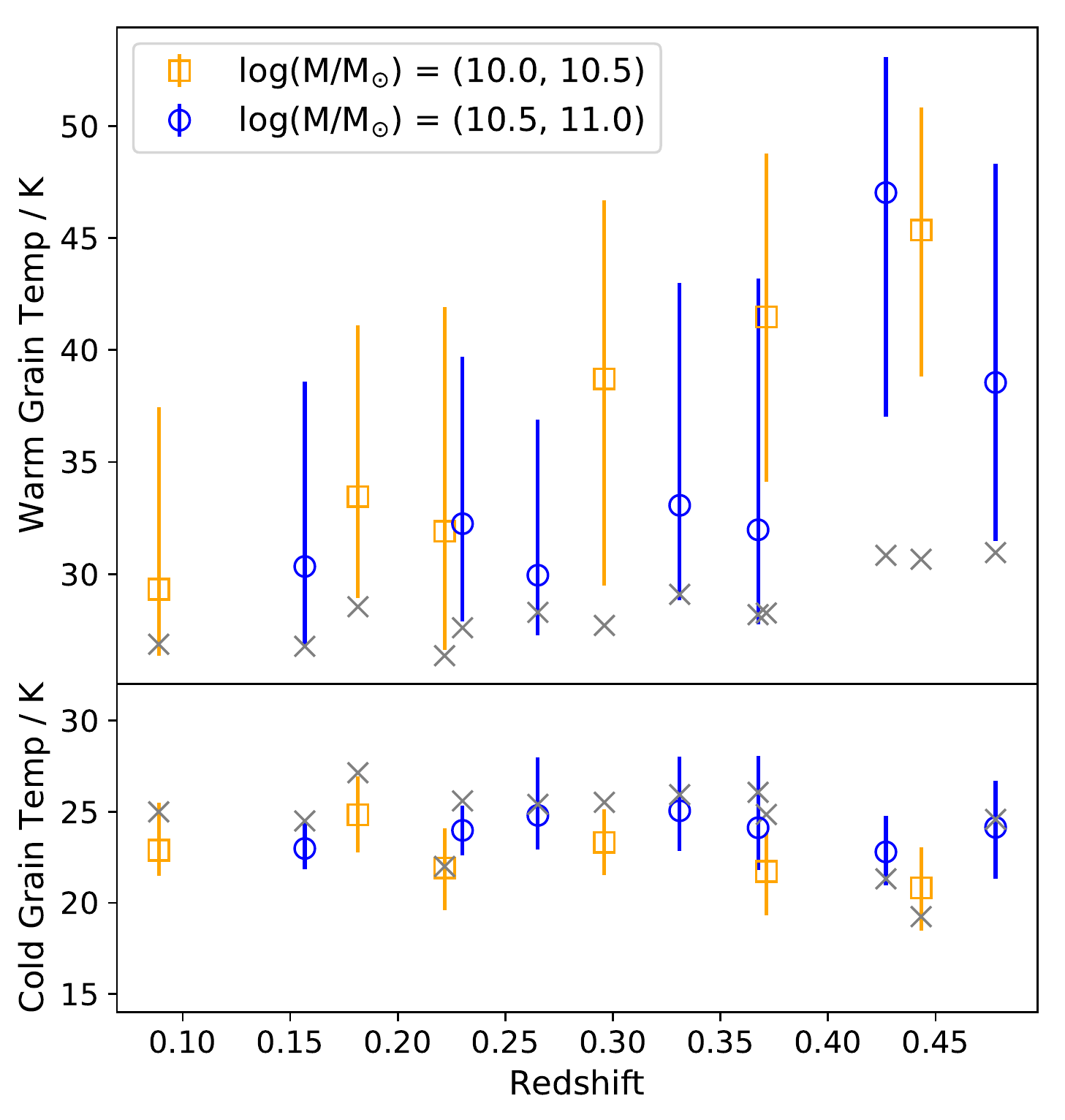}
    \caption{Same as Figure \ref{fig:cunha_temp_bpass}, but with the best fit parameters taken from fitting observations with the BC16 stellar models with a 3~Myr birth cloud age. The grey crosses overlay the median values from Figure \ref{fig:cunha_temp_bpass}.}
    \label{fig:cunha_temp_bc16}
\end{figure}

Figures~\ref{fig:cunha_temp_bpass} and~\ref{fig:cunha_temp_bc16} show the variation in the derived temperatures for the warm and cold grains (T$_W$ and T$_C$) in the dC08 model. Figure ~\ref{fig:cunha_temp_bpass} shows fits using BPASS with a 5~Myr birth cloud, while Figure~\ref{fig:cunha_temp_bc16} fits BC16 stellar models with a 3~Myr birth cloud age. In both cases, results are shown for two independent bins in stellar mass. Both stellar models are seen to have an increasing T$_W$ with redshift while T$_C$ is approximately constant over this redshift range. The cold grain temperatures do not vary significantly between BPASS and BC16 fits, but the same is not as true for the warm grain temperature. At low redshifts, the derived T$_W$ for the two stellar models is consistent within its uncertainties, but this is no longer true at the higher redshifts. 

These temperatures are constrained in the fit almost entirely by data from the {\em Herschel} space telescope. Herschel photometric data was taken by the COSMOS programme at 100, 160, 250 and 350~$\mu$m in the observer frame. These bands straddle the peak of the dust curve for the cold dust component, since a cold grain temperature of T$_C$~=~23~K has a greybody emission spectrum peaking at 126~$\mu$m. By $z=0.5$, this has redshifted to 189~$\mu$m, resulting in all four wavebands providing a strong constraint on the shape of the cool dust emission curve.

For a warm dust grain temperature between T$_W$~=~30 and 45~K the emission curve would peak between 97 and 64~$\mu$m in the rest frame. Thus the warm grains are only strongly constrained by the 100~$\mu$m observation at $z=0$, and unless the temperature is at the lower end of the potential range, this remains true at $z=0.5$. If the temperature difference between the two grain components is small, the uncertainties on the relative contribution of each will be large, since the two dust emission curves blend. This is the scenario favoured in the BPASS stellar population model fits, in which the warm dust temperature does not exceed 31\,K. Fits using the BC16 stellar models favour a much warmer dust component of 50\,K, but its precise temperature remains poorly constrained since only one data point substantially influences the fit. This results in the increased uncertainty range in the inferred SEDs shown in Figure~\ref{fig:model_fits_grid_group6} between 24 and 100~$\mu$m. Further observations in this challenging wavelength regime are required to better constrain the derived parameter space for a given input stellar model prescription.

Only having one photometric observation allows for more variation in the derived value of a parameter. In order to match to this observation point, a blackbody spectrum can modify its shape and energy by changing the temperature of the model, or by increasing the intensity of the starlight radiation spectrum. Due to the harder ionising spectrum of BPASS, more energy is being re-radiated by the dust than in the BC16 models, and so a lower temperature is required to match the observed flux of the single observation point. Conversely, for the BC16 models, the blackbody spectrum requires a higher temperature to produce the same amount of flux. This leads to the higher median in the temperature for the BC16 than the BPASS models. However, with more observations, this discrepancy may be reduced.


\subsection{Implications for the Cosmic Star Formation Rate Density History}

One of the most important and widely calculated properties of galaxy stellar populations to arise from SED fitting analyses is the shape and calibration of the cosmic star formation rate density (CSFRD) evolution with redshift. This is determined through the analysis of volume-complete or corrected galaxy surveys, in which the instantaneous star formation rate of each galaxy is inferred from the luminosity of emission lines, SED template fitting, or from the thermal infrared luminosity, or a combination of these methods. Each of these SFR conversions is fundamentally calibrated against a stellar population synthesis model which relates ionizing photon production rate to star formation rate given stated assumptions. A long standing dilemma arising from these analyses is that the integral of the CSFRD should recover the in-situ stellar mass density at a given redshift. As \citet{2014ARA&A..52..415M} noted, compilations of observational data now predict a relatively smooth CSFRD which can be parameterised as a function of redshift, but which overpredicts the cosmic stellar mass density by $\sim0.2$\,dex at low to intermediate redshifts. The mass density itself is also derived primarily from SED template fitting.

\citet{2019MNRAS.490.5359W} evaluated the impact of BPASS binary population synthesis models on these quantities, using the same v2.2.1 models used here, but adopting a very simplified dust emission model. They found that the effect of BPASS binary population synthesis calibration is to reduce the normalisation of the CSFRD by $\sim0.16$\,dex, while the mass density is reduced by $\sim0.05$\,dex. These recalibrations were based on stellar population models derived from simple assumptions of a constant SFR over timescales of $>$100\,Myrs.

The results in Section \ref{sec:impact} imply that the star formation rates of the infrared luminous galaxies that dominate star formation at $0<z<0.5$ are actually $\sim0.31\pm0.18$\,dex lower using BPASS fitting than with single-star stellar population synthesis models.  Considering only the best-characterised galaxies increases the significance of this offset, with galaxies in the range $10.5$<log(M/M$_\odot$)<$11.0$ showing an offset in SFR of $0.33\pm0.06$\,dex, although in our highest mass bin the offset reduces in both size and significance. 

Figure~\ref{fig:stel_param_comp} also shows that the offset has a slight dependence on redshift, with BPASS-derived SFRs coming closer to those derived by the COSMOS team with increasing redshift. There could be a number of reasons for this trend. The typical star formation rate of the galaxies in each subsample increases with redshift, meaning that the stellar SED becomes increasingly dominated by the youngest stellar population. This acts to diminish the uncertainties associated with the adopted star formation history, reducing a simple parameterisation of rising or falling star formation rates to essentially a short lived burst. At lower redshifts, the typical star formation rate is also lower, leading to a bigger uncertainty in establishing the contribution to the rest-frame optical and ultraviolet SED from slightly older underlying stellar populations. These cannot securely be accommodated in a simple parametric star formation history without added uncertainty. These slightly older ($\sim$100\,Myr - 1\,Gyr) populations are also the most likely to show the impact of binary stellar evolution processes extending the luminous lifetimes of moderate mass stars. Our analysis suggests lower stellar masses using BPASS by $0.14\pm0.02$\,dex (or $0.13\pm0.02$\,dex in the mass bin above). These results are broadly consistent with those of \citet{2019MNRAS.490.5359W}, with the extra offset in SFR suggested by this work balanced by a similarly increased offset in stellar mass. Thus the results here suggest that (if these results are representative of those at higher redshift) the local CSFRD integral and mass density can be reconciled, as \citet{2019MNRAS.490.5359W} suggested, as long as binary population synthesis and careful dust modelling is considered.


\section{Limitations and Uncertainties} \label{lims}


\subsection{Assumptions in the Stellar Component} \label{stel_assump}

The metallicity assumed for each galaxy was Solar ($Z=0.020$). While there is dispersion in the metal enrichment of galaxies, the local galaxy population has been shown to have typical metallicities which are approximately Solar \citep{2005MNRAS.362...41G, 2021MNRAS.502.4457G}. The well-known mass-metallicity relationship at $z\sim0$ would suggest that our lowest mass galaxy composites might be better represented by a lower metallicity stellar population. However, we do not see any evidence for poorer fits amongst these low mass bins than their higher mass counterparts. It is likely that, while fitting the stellar population with lower metallicities might slightly improve modelling of the optical region, small differences in the stellar prescription would not have affected the dust emission fitting which often dominates the uncertainties. 

The stellar population models incorporated a simple prescription for the emission of reprocessed ionizing radiation by nebular gas. The nebular gas density and geometry was fixed, as was its ionization parameter which was held constant at $\log(U_\mathrm{neb})=-3$. This value has been shown to be typical of nearby galaxies when using the BC03 and BC16 models \citep{2016MNRAS.462.1415C, 2017MNRAS.468.2140C, 2017MNRAS.470.3532V}. However, we noted in Section~\ref{sec:bpass} that the harder ionizing spectra of BPASS models at a given age may require a larger nebular ionization parameter of $\log(U_\mathrm{neb})=-1.5$ or $-1.0$ \citep{2018MNRAS.477..904X}. Fully constraining the nebular gas properties would add an additional four or more free parameters to our fitting procedure, and we chose to fix these in order to provide a direct comparison between the assumed stellar and dust models. We note that inclusion of the higher ionization potential of binary stellar populations might have increased the emission in the near-infrared part of the spectrum, producing a slightly better fit to the data. However the differences in likelihoods and thus BICs between models is dominated by the dust emission curve, rather than the few affected near-infrared data points.

We also assumed a simplified, parametric star formation history for the galaxy composites. These were selected to include relatively massive, infrared luminous galaxies and so included both an old underlying and a recently star-forming population. The age of the young stellar population is difficult to derive \citep{2014prpl.conf..219S}. A burst-like stellar population ages rapidly, and its UV (and hence IR) emission is strongly sensitive to age. An assumed single age burst at 1\,Myr would produce very different emission to the same burst seen at 10 Myr. The young population was therefore assumed to have a composite stellar population modelled as delayed-tau SFH with an age of 5~Myr. Since massive stars have lifetimes <~10~Myr, such a composite population around an age of 5~Myr is representative of the majority of UV emission. If we were concerned with the detailed star formation histories of individual galaxies, a more robust result might be obtained by fitting a fully non-parametric star formation history, but this would add substantially to the number of free parameters and degeneracies in the best fitting models. Since the data being fitted here are typically composites of $\sim$15-50 galaxies, we expect slight differences in the detailed star formation histories to cancel one another out, and hence that a simple parameterisation will provide a reasonable fit to the composites, enabling a more straight forward comparison between different stellar and dust model assumptions.

Finally, we note that the different model combinations required different assumptions regarding the existence and impact of dense, highly extincting stellar birth clouds. The composition and grain characteristics of these birth clouds represent a key assumption. Birth cloud dust is known to cause more extinction than ISM dust, leading, for example, to the higher extinction applied to nebular lines than the continuum in the \citet{2000ApJ...533..682C} extinction law. This was accounted for, in all of the dust models, with an $\eta$ parameter which instructs the \bagpipes\ fitting software of how much additional extinction to apply to young stellar populations. However, there is no general consensus on the quantitative difference between the two dust components. While some previous work has found large differences in the strength of birth cloud extinction \citep[e.g.][]{2000ApJ...533..682C, 2009ApJ...706.1364F, 2013ApJ...777L...8K, 2013ApJ...771...62K, 2014ApJ...788...86P, 2015A&A...582A..80T}, others have found values much closer to the ISM extinction \citep[e.g.][]{2006ApJ...647..128E, 2007ApJ...670..156D, 2010ApJ...712.1070R, 2012ApJ...744..154R, 2013ApJ...777L...8K, 2015ApJ...804..149S, 2016A&A...586A..83P}. The original work by \citet{2008MNRAS.388.1595D}, based on the prescription of \citet{2000ApJ...539..718C}, accommodated the birth cloud by adopting a much steeper dust attention law for the youngest stars. This led to an integrated excess in the reprocessed energy of a factor of $\sim2$, which is very similar to the 0.4 magnitude additional extinction applied by \citet{2000ApJ...533..682C} to nebular regions in their extinction law. We thus adopted a scaling factor of 2.0.

In this analysis we only tested the BPASS binary stellar models, but other binary models are starting to appear in the literature \citep[e.g.][]{2018A&A...615A..78G, 2019A&A...629A.134G, 2022arXiv220205892F}. We expect that, in general terms, the evolution of binary populations will be similar between these models, as also found by \citet{2019A&A...629A.134G}. However, we note that the detailed implementation of binary phenomena such as mass transfer, common envelope evolution, supernova kicks and the subsequent disruption of binaries, are handled differently in different codes. Thus while a population of stripped helium stars will emerge in any binary population synthesis, which dominate the broad trends in the population evolution, the precise timing and contribution to the total luminosity will depend on the details of the formalism adopted.


\subsection{Limitations to the Dust Emission Models}

All of the dust emission models have built-in assumptions to simplify them. These assumptions affect their ability to fit to observations in ways which we outline here.

The dC08 generated emission models are a combination of four emission components, each of which have their own parameterisation. While dC08 fit to observations of high Galactic latitude dust emission in the Milky Way to constrain parameters associated with the ambient ISM component, this is not strictly valid for galaxies which differ substantially in star formation history and metallicity to the Milky Way. Hence we leave them all as adjustable parameters. This means our empirical dust emission model has 7 free parameters, which is equal to the number of photometric observations available to constrain the infrared region of the SED. Thus the dust models are underconstrained, leading to a wider range of dC08 models which can map to observations when compared to other dust models considered here. This was highlighted in Figure~\ref{fig:model_fits_grid_group6}. To relieve pressure on the fit, we fixed some parameters when generating the empirical model, such as requiring the birth cloud and ISM component to use the same warm grain temperature. In practice the birth cloud may have hotter dust due to closer proximity to stellar irradiation.

The model of DL07 made assumptions about how to incorporate different dust emission due to close proximity to stellar radiation. For this, they assumed that some dust grains are illuminated by a minimum (ISM-like) radiation field while the rest receive increased (birth cloud-like) radiation following a power-law slope distribution up to a fixed maximum radiation field. An additional parameter allows the relative contribution from these two components to be determined. In doing this, it means that all stellar energy attenuated is combined into a total dust energy, before being redistributed by thermal processes. Thus, when the prescription varies for young stellar populations, i.e. increased ionising flux from binary stars, there is no specific dust component which can absorb these variations. Instead, the whole dust emission spectrum is modified, causing problems when fitting with certain stellar population synthesis models. This lack of a clearly-defined birth cloud component may make the dust emission prescription of DL07 too simplistic to be applied when considering BPASS stellar population models.

D20 dust emission models were generated by illuminating the dust with different aged stellar populations to determine the evolution of the dust emission shape. We adopted dust models generated by radiation from a 3~Myr and 1~Gyr stellar population to simulate the birth cloud and ISM components respectively. However, a correct prescription should self-consistently generate dust models for all available stellar population ages (41 timesteps in the case of BPASS models) and then have the same stellar SFH applied to work out the contribution from each dust emission model. Due to a limited number of models in the publicly available grid generated with differently aged stellar spectra, this was not feasible. However, the use of only two models is a reasonable approximation as the young stellar population in our star formation history prescription had a fixed age of 5~Myr and the older population was allowed to vary over ages of a few Gyr, at which ages the stellar population changes only slowly. As a result stellar populations in this age range generate approximately the same ultraviolet and blue-optical radiation fields as that which generated the dust models used.


\section{Conclusion} \label{conc}

In this paper we have presented an analysis of the joint impact of assumed stellar irradiation spectrum and dust emission prescription on the interpretation of galaxy properties including age, extinction, mass, SFR and dust temperatures. We have used a sample of infrared luminous galaxies at $0<z<0.5$ drawn from the COSMOS survey \citep{2007ApJS..172....1S, 2016ApJS..224...24L} as a demonstration and evaluation tool. To allow a direct comparison between combinations of stellar and dust model assumptions, we fit these with a simple parameterised star formation history, using the SED fitting tool \bagpipes\ \citep{2018MNRAS.480.4379C}. We summarise our principle conclusions as follows:

\begin{enumerate}
    \item When fitting a multiwavelength galaxy SED extending to sub-millimeter wavelengths, the derived properties of the stellar population depends not only on the stellar population synthesis models deployed, but also on the dust emission prescription.
    \item Conversely, the derived dust properties in such a fit, depend not only on the dust emission curve but also the stellar population. This affects parameters such as the dust temperature (see Figure \ref{fig:cunha_temp_bc16}).
    \item The fitting procedure adopted here, results in a consistently younger inferred stellar population for a given set of galaxy photometry than was found by the COSMOS2015 programme. All galaxy mass and redshift bins are broadly consistent with \bagpipes\ derived stellar population ages in the range 0.5-2\,Gyr.
    \item Stellar masses inferred using the \bagpipes\ algorithm, when the FIR is fit simultaneously with optical and ultraviolet fluxes, are consistent with those derived by the COSMOS team with a typical scatter of 0.2 dex. However, masses inferred using BC16 are $0.14\pm0.02$\,dex higher than those inferred using the BPASS stellar population models. This arises due to the extra ultraviolet flux available in the BPASS models, rather than differences in the stellar models at long wavelengths.
    \item Binary population synthesis corrections to the cosmic star formation rate density calibration and the stellar mass densities estimated from SED fitting may help to reconcile the discrepancy between these empirical quantities.
    \item The simple star formation history prescription adopted here favours lower dust extinctions than found by the COSMOS2015 fitting algorithm, but the offset depends on stellar population synthesis models. BPASS models produce more ultraviolet luminosity for a given mass and therefore require extinctions lower by E(B-V)~=~$0.064\pm0.022$ than the BC16 models given the same prescription.
    \item All of the stellar and dust model combinations considered here provide reasonable fits to the data. However, the lowest values of the Bayesian Information Criterion are obtained using the Bruzual and Charlot models together with the DL07 dust model grid which was informed by its radiative spectrum. However, the lack of binary stellar evolution pathways in this population synthesis raises questions over its reliability in young, massive star dominated populations. The BPASS models typically show a higher (and therefore weaker) Bayesian Information Criterion value. The best fits to the data were obtained using the D20 dust emission models which were specifically irradiated by BPASS spectra, and therefore consider dust reemission self-consistently. This emphasises the interdependence of dust and stellar population synthesis models.
    \item When fitting with binary stellar populations, the longer ultraviolet-luminous epoch after stellar birth renders energy balance calculations sensitive to the dispersal timescale of the dusty stellar birth cloud. Optimal fits were obtained with a dispersal timescale of at least 5\,Myrs. By contrast, irradiation by BC03 and BC16 stellar models shows no dependence on the birth cloud lifetime beyond 3\,Myrs.
    \item Fits using the dC08 dust prescription when the BC16 stellar model is used to irradiate the cloud require warm grain temperatures between 30 and 50\,K, with a weak indication of rising temperature with redshift. By contrast for irradiation with BPASS, the same model requires dust temperatures in a narrow range of 26-31\,K, as a result of the stronger ultraviolet emission.
    
\end{enumerate}

As a consequence of these results, we strongly recommend that the irradiating spectrum be considered as an integral element of dust emission modelling. It is important that further investigations be made into the impact of stellar assumptions on the reprocessing of energy by dust. The analysis here has used a simple star formation history to allow direct comparisons over a large model grid. It is possible that the differences observed between model combinations would be reduced were more flexibility be permitted in the star formation histories. However in this case the uncertainty would be transferred to the evolutionary history and interpretation of the galaxies themselves. 

We note that this study has been limited to the intensely star forming galaxies observable through the far-infrared in the very local Universe. New facilities, notably the James Webb Space Telescope, will soon be providing new constraints on the mid-infrared portion of the dust emission curve in these local galaxies, and beginning to place constraints to somewhat higher redshift. Meanwhile observations with the Atacama Large Millimeter Array (ALMA) are beginning to build a catalogue of typical star forming galaxies at Cosmic Noon and beyond for which multiple continuum data points are now available in the mid- to far-infrared. 

A very recent analysis by \citet{2022arXiv220302059B} for example, has used the {\sc CIGALE} SED fitting code to explore the properties of a large sample of infrared-detected high redshift galaxies identified in the ALPINE survey. As the results here have demonstrated, any such analysis is subject to systematic offsets due to the model grid used, and going into the high redshift regime, the dust emission and stellar emission spectra may both vary due to differences in the chemical composition, structure and stellar population parameters, which is a key area for further investigation.


\section*{Acknowledgements}

ERS and GTJ acknowledge support from the UK Science and Technology Facilities Council (STFC) through consolidated grant ST/T000406/1 and a doctoral studentship respectively. ACC would like to thank the Leverhulme Trust for their support via the Leverhulme Early Career Fellowship scheme. We thank members of the BPASS team, past and present, for helpful insights. This work made use of the University of Warwick Scientific Computing Research Technology Platform (SCRTP) and Astropy\footnote{\url{https://www.astropy.org/}}, a community-developed core Python package for Astronomy \citep{astropy:2013,astropy:2018}. 

\section*{Data Availability}
All photometric and modelling data underlying these results are publicly available. Specific results of fits will be made available via the BPASS website at \url{www.warwick.ac.uk/bpass} or \url{bpass.auckland.ac.nz}, or by reasonable request to the first author.


\bibliographystyle{mnras}
\bibliography{paper}



\clearpage

\appendix

\section{Model fits}

\begin{figure*}
	\includegraphics[width=\textwidth]{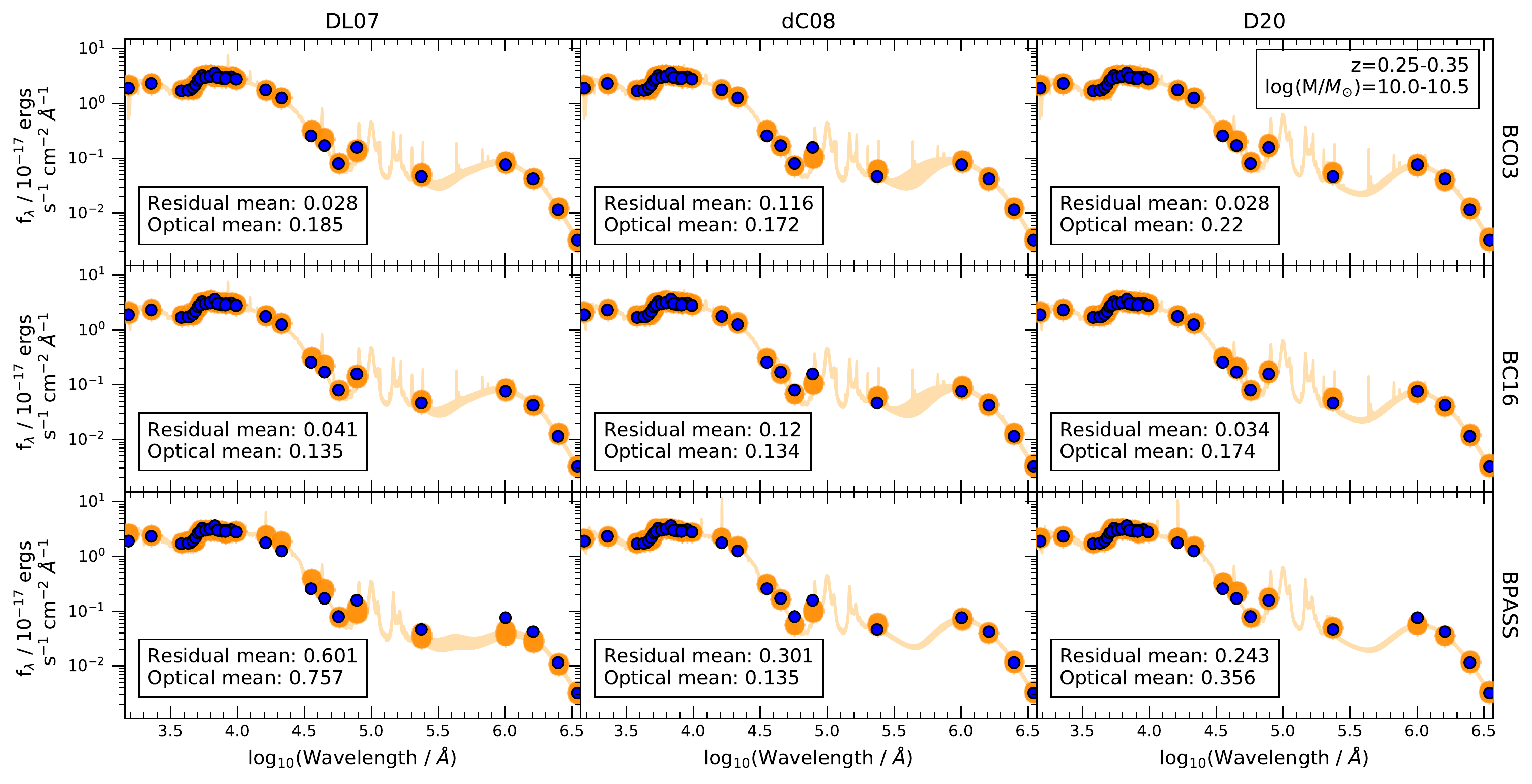}
    \caption{Best-fit models from different stellar and dust emission model combinations fitted to stacked galaxy data from the bin at z=0.25-0.35 and log(M/$M_{\odot}$)=10.0-10.5 with a 5 Myr birth cloud age. The rows contain the different stellar models of BC03 (top), BC16 (middle) and BPASS (bottom), each combined with the different dust emission models of DL07 (left column), dC08 (middle column) and D20 (right column). Each combination has plotted the best-fit spectrum (orange line) along with the expected photometric flux in each filter for that spectrum (orange dots), overlaid with the observation data in each filter (blue dots). Errors have been included for the observational data but are too small to be seen, while the thickness of the orange line represents the uncertainty in the model. Each plot also contains the mean normalised residual when averaging over all photometric filters and the optical normalised residual when averaging only over the optical and near-infrared filters (i.e. those where the stellar spectrum dominates). Note that the models have been redshifted to the mean redshift of the bin (z~=~0.298).}
    \label{fig:model_fits_all_grid_group6}
\end{figure*}

Figure~\ref{fig:model_fits_all_grid_group6} shows the fits to the galaxy observations in the bin z=0.25-0.35 and log(M/$M_{\odot}$)=10.0-10.5 for all model combinations, assuming a birth cloud age of 5 Myr. Each plot contains the mean residual when averaging over all observations and the mean optical residual when averaging over only the optical and near-IR filter observations (i.e. those where the stellar spectrum dominates). The figure highlights that all model combinations can reproduce the observations with reasonable agreement, but the BPASS stellar model with DL07 dust emission model is the worst fit, with much larger residuals, especially in the optical photometric points to counter-act the poor fitting in other regions of the SED.

\begin{figure*}
	\includegraphics[width=\textwidth]{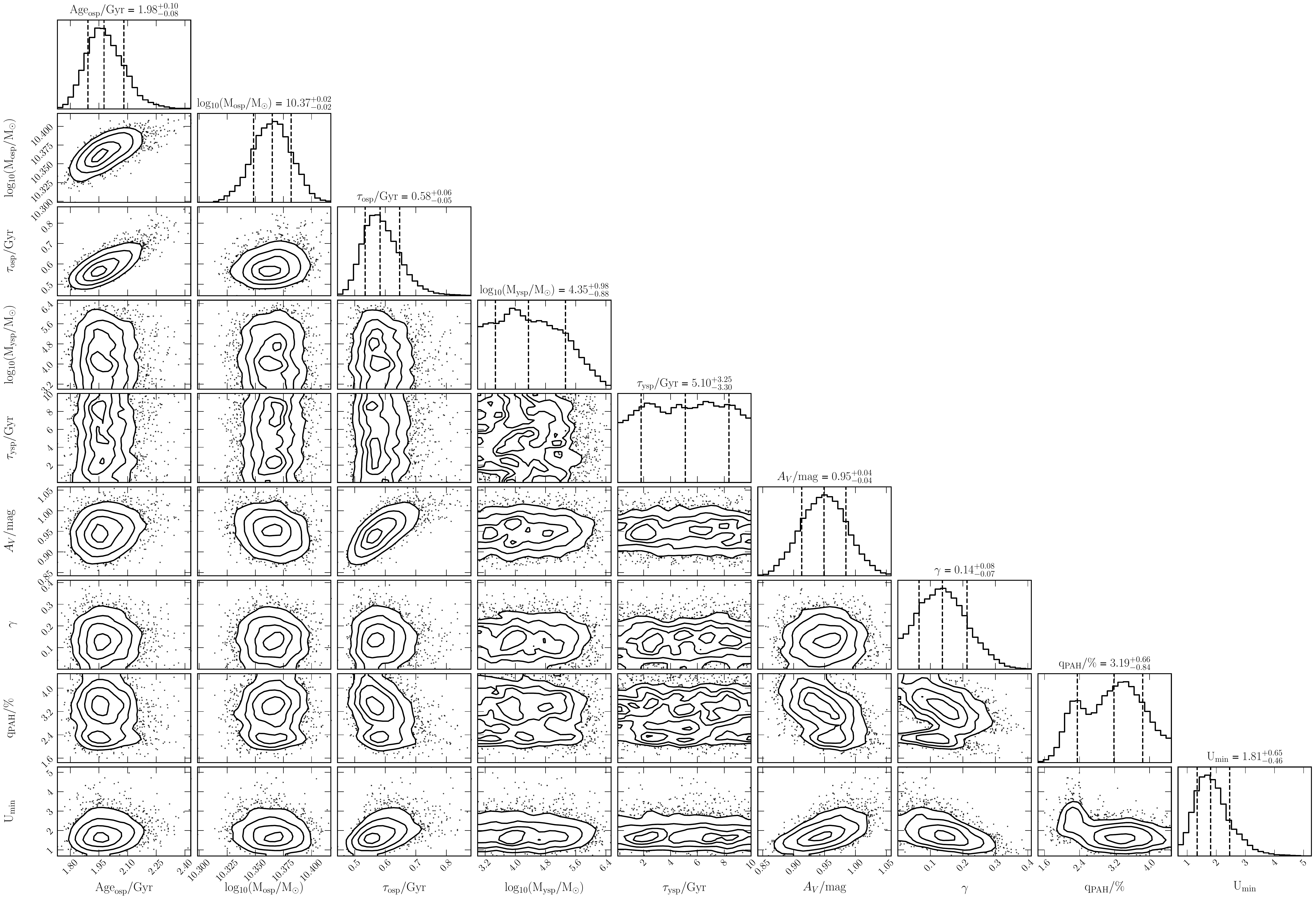}
    \caption{An example of a posterior probability distribution corner plot from \bagpipes\ fitting of an individual stacked SED from the bin at z=0.25-0.35 and log(M/$M_{\odot}$)=10.0-10.5 using the BPASS stellar and DL07 dust emission model combination. The first five columns/rows are the stellar parameters, with the first three related to the age, mass and $\tau$-parameter of the SFH associated with the older stellar population (osp), while the next two relate to the mass and $\tau$-parameter associated with the young stellar population (ysp) of age 5~Myr. The last four columns/rows are the dust parameters, starting with the extinction, $A_V$ in magnitudes, followed by the three DL07 dust emission model parameters of $\gamma$, $\mathrm{q_{PAH}}$, and $\mathrm{U_{min}}$. The top panel in each column shows the probability distribution of that parameter, with the 16$\%$, 50$\%$ and 84$\%$ quartiles shown as the dashed lines. All other panels shows the two-dimensional probability distribution of the overlapping parameters in that row and column. This demonstrates the degeneracies which lead to thickening of the best-fitting spectrum in Figure \ref{fig:model_fits_grid_group6}.}
    \label{fig:model_corner_plot}
\end{figure*}

In Figure \ref{fig:model_corner_plot} we show an example of a posterior probability distribution corner plot from \bagpipes\ fitting of an individual stacked SED from the bin at z=0.25-0.35 and log(M/$M_{\odot}$)=10.0-10.5 using the BPASS stellar and DL07 dust emission model combination. This demonstrates the dust temperature-related parameter degeneracies which lead to thickening of the best-fitting spectrum in Figure \ref{fig:model_fits_grid_group6}. The young stellar population parameters of mass and delayed-tau parameter of the SFH are the constrained the poorest due to a combination of the young population not being dominant enough to be separated from the older population and the dust emission model having no separate birth cloud component, in which the amount of flux from the young stellar population is a key component.

\section{BICs}

\begin{figure}
	\includegraphics[width=\columnwidth]{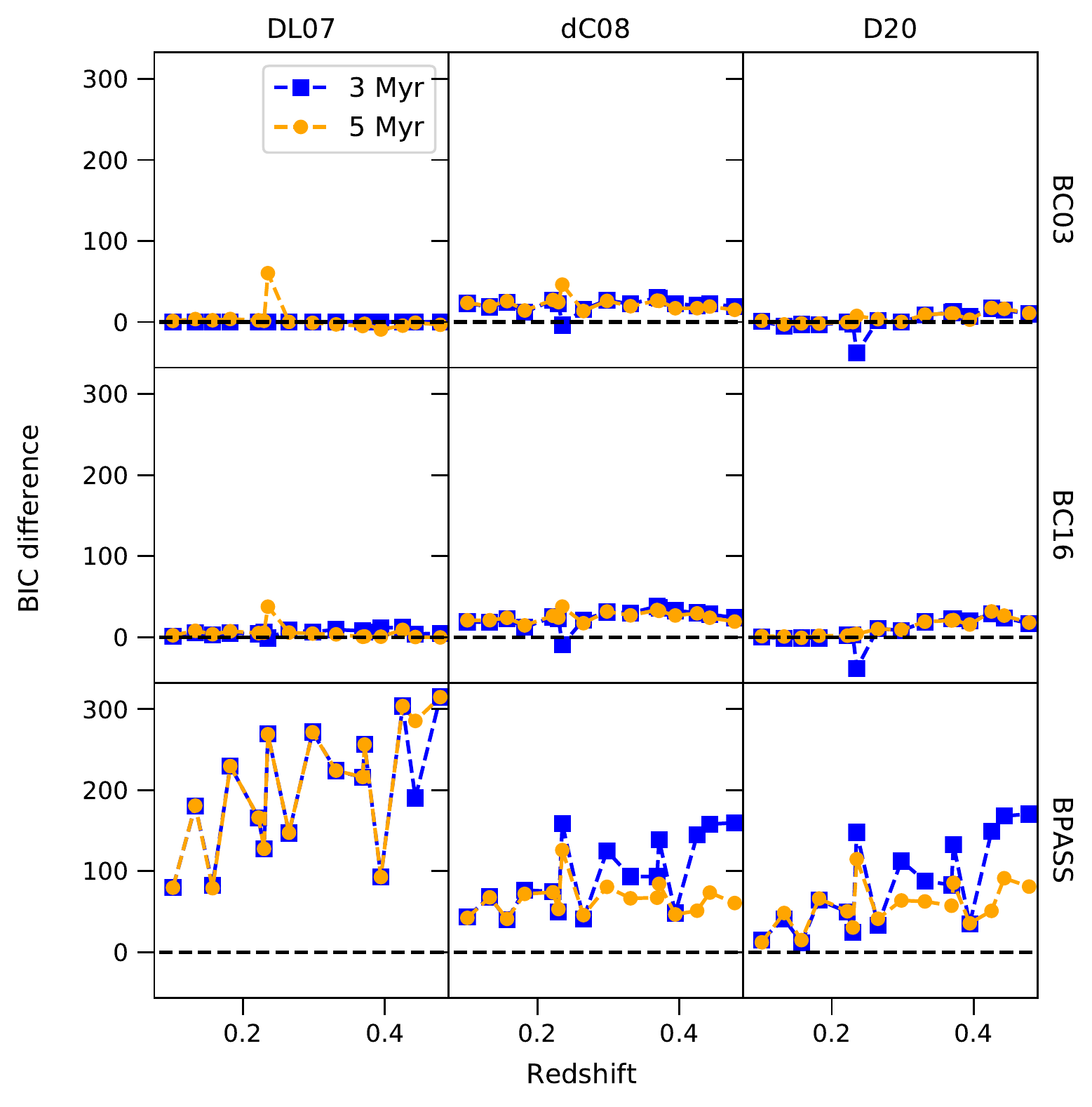}
    \caption{BIC values for all model combinations, calculated as the difference to the BC03 and DL07 model combination with a 3~Myr birth cloud age. The rows contain the different stellar models of BC03 (top), BC16 (middle) and BPASS (bottom), each combined with the different dust emission models of DL07 (left column), dC08 (middle column) and D20 (right column). The blue squares show the resulting BICs for a 3~Myr birth cloud age while the orange circles are for a 5~Myr birth cloud age. Lower BIC values and hence lower BIC differences indicate that is the preferred model.}
    \label{fig:all_likes}
\end{figure}

Figure~\ref{fig:all_likes} shows the BIC values for all model combinations and all stacked galaxy samples, for both a 3 and 5~Myr stellar birth cloud age. These are shown as the difference to the BC03 and DL07 model combination with a 3~Myr birth cloud age. A lower BIC value and hence a lower BIC difference indicates that model is the preferred one out of the two being compared. There is little difference between the BIC values for the BC03 and BC16 models, meaning that the slight prescription changes to the single evolution code has not altered the fitting performance. The dC08 dust emission models have similar trends to the D20 models but lie at slightly higher BIC values. This is due to the first term of the Bayesian Information Criterion which takes into account the number of parameters in the model, k, and increases as the number of parameters increases. For the DL07, dC08 and D20 dust models, k~=~9, 13, and 10 respectively. Therefore, since the dC08 model has a larger number of parameters, the BIC value increases creating this slight offset compared to the other models.

\bsp	
\label{lastpage}
\end{document}